\documentclass[aps,prd,preprintnumbers,showpacs,nofootinbibn]{revtex4}

\usepackage{multirow}
\usepackage[english]{babel}
\usepackage{amsmath}
\usepackage{amssymb}
\usepackage{epsfig}
\usepackage{graphics,psfrag,rotating}
\usepackage{graphicx}
\usepackage{dcolumn}
\usepackage{bm}
\usepackage{setspace}
\usepackage[T1]{fontenc}
\usepackage[utf8]{inputenc}
\usepackage{babel}
\usepackage[font=small,labelfont=bf,tableposition=bottom,labelformat=empty]{caption}

\begin{document}
\newcommand{\Od}{{\cal O}}
\newcommand{\lsim}   {\mathrel{\mathop{\kern 0pt \rlap
  {\raise.2ex\hbox{$<$}}}
  \lower.9ex\hbox{\kern-.190em $\sim$}}}
\newcommand{\gsim}   {\mathrel{\mathop{\kern 0pt \rlap
  {\raise.2ex\hbox{$>$}}}
  \lower.9ex\hbox{\kern-.190em $\sim$}}}


\title{Spectral Study of the HESS J1745-290 Gamma-Ray Source as Dark Matter Signal} 

\author{J. A. R. Cembranos\footnote{E-mail:cembra@fis.ucm.es}, V. Gammaldi\footnote{E-mail:vivigamm@pas.ucm.es}, and A.\,L.\,Maroto\footnote{E-mail:maroto@fis.ucm.es}}

\affiliation{Departamento de  F\'{\i}sica Te\'orica I, Universidad Complutense de Madrid, E-28040 Madrid, Spain}%
\date{\today}

  \begin{abstract}
                We study the main spectral features of the gamma-ray fluxes observed by the High Energy Stereoscopic System (HESS) from the J1745-290 Galactic Center source during the years 2004, 2005 and 2006. In particular, we show that these data are well fitted as the secondary gamma-rays photons generated from dark matter  annihilating into Standard Model particles in combination with a simple power law background. We present explicit analyses for annihilation in a single standard model particle-antiparticle pair. In this case, the best fits are obtained for the $u\bar u$ and $d\bar d$ quark channels and for the $W^+W^-$ and $ZZ$ gauge bosons, with background spectral index compatible with the Fermi-Large Area Telescope (LAT) data from the same region. The fits return a heavy WIMP, with a mass above $\sim10$ TeV, but well below the unitarity limit for thermal relic annihilation.
  \end{abstract}


\maketitle

\section{Introduction}
\label{intro}

Numerous and different  data collected during the last years have established the
Standard Cosmological Model as a simple framework showing a remarkable
agreement with observations. This model is based on Einstein's General Relativity
and a homogeneous and isotropic {\it ansatz} for the metric. However,
in addition to ordinary matter, two new elements need to be added:
a cosmological constant or other form of dark energy to account for the late time
acceleration of the universe, and dark matter (DM) to explain the formation
and dynamics of cosmic structures. However, despite the multiple efforts, the fundamental
nature of DM remains still as an open problem. There are strong astrophysical evidences for
DM from galactic to cosmological scales, but the interactions with ordinary matter have not
been probed beyond gravitational effects. In this sense, both direct and indirect DM searches
are fundamental to explore particle models of DM.

In the framework of indirect searches, the observation of gamma-ray fluxes from astrophysical sources
represents one of the main approaches to the DM puzzle. If DM annihilates into SM particles, the
subsequent chains of decay and hadronization of unstable products produce gamma-ray photons generically.
The observation of this signal is highly affected by astrophysical uncertainties in the gamma-ray backgrounds
and in the DM densities and distribution. On the other hand, depending on both astrophysical and particle 
physics models of DM, the signature could be distinguishable from the background. Appealing targets for 
gamma-ray observations of annihilating DM are mainly selected according to the abundance of DM in the source 
and their distance. Galaxy clusters, dwarfs spheroidal
galaxies (dSph) or the galactic center (GC) of the Milky Way are traditional targets of interest. Galaxy clusters are very rich in
DM, but they are very distant objects. DSphs are also characterized by high DM densities but at much shorter distances.
In any case, despite their proximity, observations of dSphs have not been able to detect any gamma-ray flux signature over the background
so far \cite{SEGUE, FerdSp}.
On the other hand, the GC represents a very close target, but the complexity of the region, due to the
high concentration of sources, makes the analysis quite involved.

In this work, we will focus on the analysis of very high energy (VHE) gamma-rays coming from the GC.
Different observations from the GC have been reported by several collaborations such as CANGAROO \cite{CANG}, VERITAS \cite{VER},
HESS  \cite{Aha, HESS}, MAGIC \cite{MAG} and Fermi-LAT  \cite{Vitale, ferm}. In particular, we will analyze the data collected by the
HESS collaboration during the years 2004, 2005, and 2006 associated with the HESS J1745-290 GC source \cite{HESS}.
The interpretation of these fluxes as DM signal has been widely discussed in the literature from the very early days of the publication
of the observed data \cite{Horns:2004bk,Bergstrom1,Profumo:2005xd,DMint}. It was concluded that
the spectral features of these gamma-rays disfavored the DM origin \cite{AN, DMint}. However, in recent studies \cite{Cembranos:2012nj,Belikov:2012ty},
it has been pointed out that these fluxes are compatible with the spectral signal of DM annihilation
provided
it is complemented with a simple background. This extra source
of gamma-rays can be originated by radiative processes generated by  particle acceleration in the vicinity of Sgr A East supernova and the
supermassive black hole \cite{SgrA}.

In this work, we present a systematic study of this assumption: In Section \ref{HESS}, we show an analysis of the
source, while a brief review of the gamma-ray flux coming from annihilating DM in galactic sources is presented in Section \ref{DM}.
The fit of the HESS data with a total fitting function given by the combination of the background power law component with annihilating
DM signature is discussed in Section \ref{fit}. In Section V, we include in the analysis the
data collected by Fermi-LAT from the same region. We summarize the main results of our work and prospects for future
analyses in Section \ref{con}. Finally, an appendix provides useful details for reproducing the statistical study performed in these
analyses.

\section{HESS J1745-290 data}
\label{HESS}

DM is expected to be clumped in the central region of standard galaxies. In this sense, the central part of the Milky Way could be an important source of gamma-rays produced in the DM annihilation processes. Because of its closeness, the GC represents a very appealing target for the indirect search of DM, but the complex nature of this area makes the identification of the sources quite difficult. Several sources have been detected not only in the gamma-ray, but also in the infrared and X-ray ranges of the spectrum. The absence of variability in the collection of HESS J1745-290 data in the TeV scale, during the years 2004, 2005 and 2006, suggests that the emission mechanism and emission regions differ from those invoked in the variable infrared and X-ray emissions \cite{X}. The significance of the signal reduces to few tenths of degree \cite{HESS}, but the angular distribution of the very high energy gamma-ray emission shows the presence of an adjunctive diffuse component. The fundamental nature of this source is still unclear. These gamma-rays could have been originated by particle propagation
\cite{ferm,SgrA} in the neighborhood of the Sgr A East supernova remnant and the supermassive black hole Sgr A, both located
at the central region of our galaxy \cite{Atoyan, AN}. If it was originated by the DM distribution, the morphology of the source requires a very
compressed DM structure. In fact, it has been claimed \cite{Blumenthal, Prada:2004pi} that baryonic gas could account for similar effects when  falling
 to the central region, modifying the gravitational potential and increasing the DM density in the center
(see however \cite{Romano, Salucci:2011ee}). If this is the case, the DM annihilating fluxes are expected to be enhanced by an important factor with respect to
DM alone simulations, such as the classical NFW profile \cite{Prada:2004pi}.

\begin{figure}[!ht]
    \centering
\resizebox{11cm}{8cm}
{\includegraphics{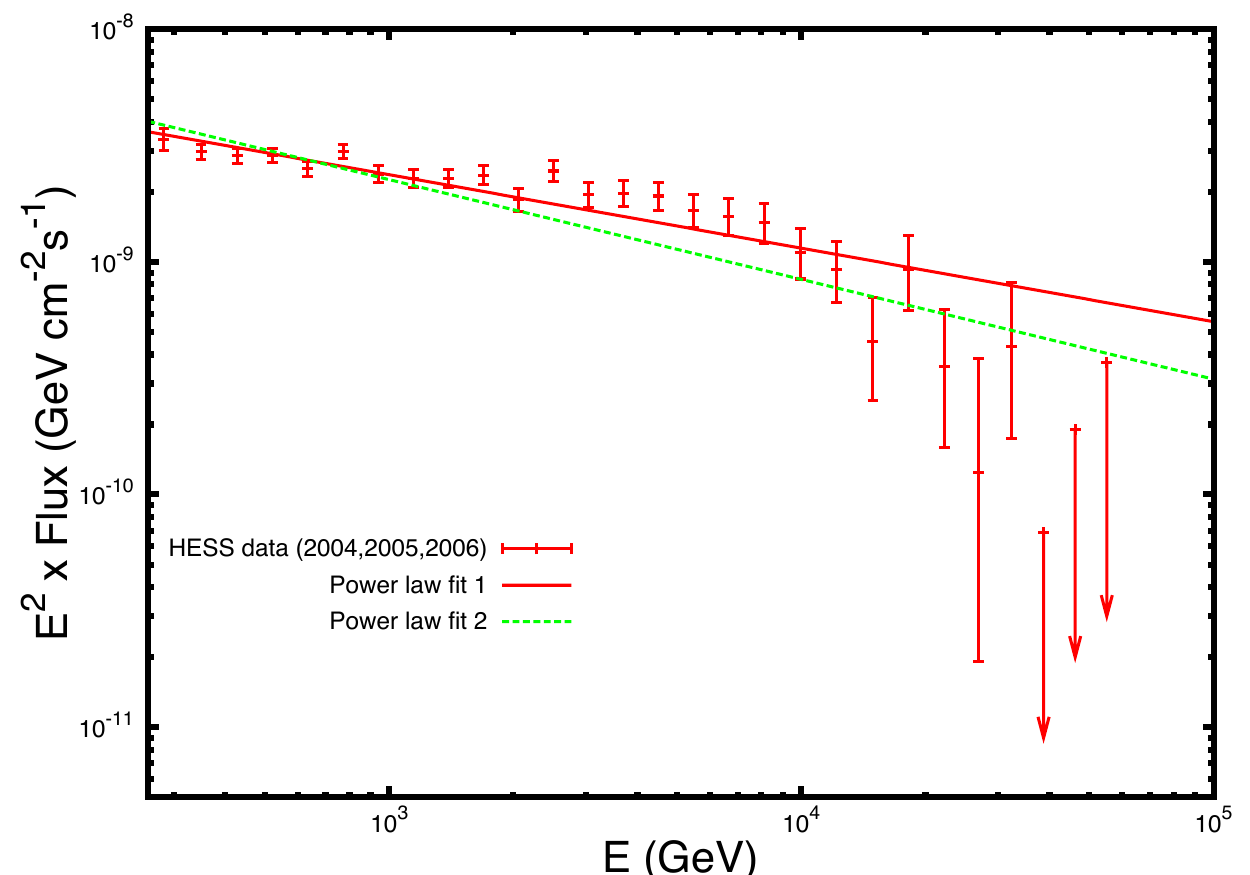}}
   \quad
  \begin{tabular}[b]{ccc}
  \hline
    \hline
    & Power law 1 &     Power law 2        \\ \hline
    \hline
    $\Phi_0$  &  $20.9 \pm 5.1$ &$43.5\pm14.8$\\ 
    $\Gamma$  & $2.32 \pm0.03$& $2.43\pm0.05$ \\
    $\chi^2/\,$dof & $2.48(57/23)$ & $5.30(137/26)$ \\
    \hline
    \hline
    &\\
    &\\
    &\\
&\\
&\\
&\\
&\\
&\\
&\\
&\\
    \end{tabular}
\caption {\footnotesize{Collection of the HESS data (2004,2005 and 2006) fitted with two simple power law background. The full line (Power law 1) shows the fit with 23 dof, without take into account the upper limit constraints on the flux given by the last three points. They are included in the Power law 2 fit with 26 dof (dotted line). The parameters of the fit can be found in the table.
Because of the poorness of both the fits ($\chi^2/dof=2.48$ with $\Gamma=2.32\pm0.03$ for Power law 1, and $\chi^2/dof=5.30$ with $\Gamma=2.43\pm0.03$ for Power law 2), they represent just an upper limit of the flux. These results are consistent with previous studies \cite{HESS, Atoyan}.}}
\label{Bgfit}
\end{figure}

\begin{figure}[!ht]
    \centering
\resizebox{11cm}{8cm}
{\includegraphics{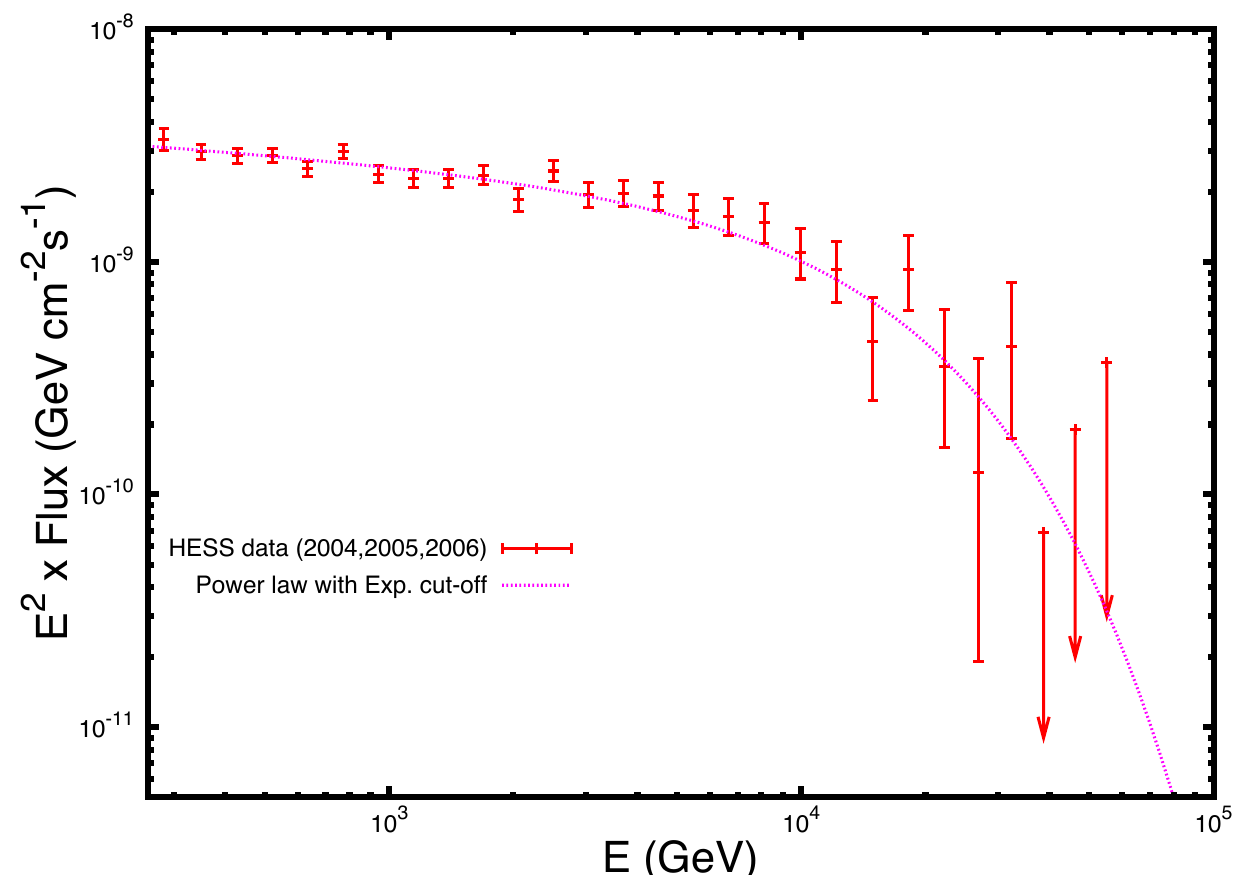}}
   \qquad
  \begin{tabular}[b]{cc}
  \hline
    \hline
   &  $Exp_{Cut-off}$ law         \\ \hline
    \hline
    $\Phi_0$  &  $6.24 \pm 1.53$ \\ 
    $\Gamma$  & $2.12 \pm0.04$ \\
    $E_{Cut}$ & $13.8\pm 2.3$ \\
    $\chi^2/\,$dof & $0.87(21/25)$\\
    \hline
    \hline
    &\\
    &\\
    &\\
&\\
&\\
&\\
&\\
&\\
&\\
&\\
    \end{tabular}
\caption {\footnotesize{
The same as in Fig. \ref{Bgfit}, but the power law is modified with an exponential energy cut-off. The signal is well fitted in this case with the results given in the attached table.}}
\label{Bgcutfit}
\end{figure}

In previous works \cite{HESS, Atoyan}, important deviations from a power law spectrum have been already reported, and a cut-off at several tens of TeVs
has been proved as a distinctive feature of the spectrum. For instance, the observed data were compared to a simple power law:
\begin{equation}
\frac{d\Phi_{Bg}}{dE}=\Phi_0\cdot \left(\frac{E}{\text{GeV}}\right)^{-\Gamma}\;,
\label{Bg1}
\end{equation}
and a power law with high energy exponential cut-off:
\begin{equation}
\frac{d\Phi_{Bg}}{dE}=\Phi_0\cdot \left(\frac{E}{\text{GeV}}\right)^{-\Gamma}\cdot e^{-\frac{E}{E_{cut}}}\;,
\label{Bg2}
\end{equation}
where $\Phi_0$ is the flux normalization, $\Gamma$ is the spectral index and $E_{cut}$ is the cut-off energy. The measured spectrum for the three-year dataset ranges from 260 GeV to 60 TeV. We have reproduced these analyses with the results that can be found in Figs. \ref{Bgfit} and \ref{Bgcutfit}. They are consistent with previous works \cite{HESS, Atoyan}, since the spectrum is well described by a power law with exponential cut-off (Fig. \ref{Bgcutfit}), while a simple power law is clearly inconsistent (Fig. \ref{Bgfit}).

\section{Gamma-rays from dark matter annihilation}
\label{DM}

If the signal observed from the GC is a combination of gamma-ray photons from annihilating DM and a background,
the total fitting function for the observed differential gamma-ray flux will be:

\begin{equation}
\frac{d\Phi_{Tot}}{dE}=\frac{d\Phi_{Bg}}{dE}+\frac{\Phi_{DM}}{dE}.
\label{gen}
\end{equation}

We will assume a simple power law background parameterized as:

\begin{equation}
\frac{d\Phi_{Bg}}{dE}=B^2\cdot \left(\frac{E}{\text{GeV}}\right)^{-\Gamma}.
\label{Bgeq}
\end{equation}

On the other hand, the differential gamma-ray flux originated from DM annihilation in galactic structures and substructures can be written
generically as:
\begin{equation}
\frac{d \Phi_{\text{DM}}}{dE} =\sum^{\text{channels}}_i \frac{\langle\sigma_i v\rangle}{2} \cdot \frac{dN^{}_{i}}{dE}
 \cdot \frac{{\Delta\Omega\,\langle J \rangle}_{\Delta\Omega}}{4 \pi M}\,,
\label{eq:totalflux}
\end{equation}
where $\langle\sigma_i v\rangle$ are the thermal averaged annihilation cross-sections of two
DM particles into SM particles (also labeled by the subindex i). We are assuming that the DM particle is its
own antiparticle. $M$ is the mass of the DM particle, and the number of photons produced in each annihilating channel
$dN_{i}/dE$, involves decays and/or hadronization of unstable products such as quarks and leptons. Because of the
non-perturbative QCD effects, the calculation of $dN_{i}/dE$ requires Monte Carlo events generators such as PYTHIA \cite{pythia}.
In our analysis, we will focus on gamma-rays coming from external bremsstrahlung and fragmentation of SM particle-antiparticle pairs produced by
DM annihilation. We will ignore DM decays, the possible production of monoenergetic photons, N-body annihilations (with $N>2$), or photons produced
from internal bremsstrahlung, that are model dependent. In particular and in order to simplify the discussion and provide useful information for a general analysis, we will consider DM annihilation into each single channel of SM particle-antiparticle pairs, i.e.

\begin{equation}
\label{singlechannel}
\frac{d\Phi^i_{\text{DM}}}{dE}= A_i^2 \cdot \frac{dN_{i}}{dE}\;,
\end{equation}
where
\begin{equation}
\label{A}
A_i^2=\frac{\langle \sigma_i v \rangle\, \Delta\Omega\, \langle J \rangle_{\Delta\Omega}}{8\pi M^2}
\end{equation}
is a new constant to be fitted together with the DM particle mass $M$, and the background parameters $B$ and
$\Gamma$. The astrophysical factor  ${\langle J \rangle}_{\Delta\Omega}$ of Eq. (\ref{eq:totalflux}) will be also indirectly fitted by means of the
parameter $A_i$. In its most general expression, it is computed in the direction $\Psi$ defined by
the line of observation towards the GC:
\begin{eqnarray}
{\langle J \rangle}_{\Delta\Omega}= \frac{1}{\Delta\Omega}\int_{\Delta\Omega}\text{d}\Omega\int_0^{l_{max}(\Psi)} \rho^2 [r(l)] dl(\Psi)\,,
\label{flux}
\end{eqnarray}
where $l$ is the distance from the Sun to any point in the halo. The radial distance $r$ is measured from the GC,
and is related to $l$ by $r^2 = l^2 + D_\odot^2 -2D_\odot l \cos \Psi$, where $D_\odot \simeq 8.5$ kpc is the distance from the Sun
to the center of the Galaxy. The distance from the Sun to the edge of the halo in the direction $\theta$ is
$l_{max} = D_\odot \cos \theta + \sqrt{r^2-D_\odot^2 \sin \theta}$.

The photon flux must be averaged over the solid angle of the detector, that is typically of order
$\Delta \Omega = 2 \pi ( 1 - \cos \Psi ) \simeq 10^{-5}$ for detectors with sensitivities in the TeV
energy scale,
such as the HESS Cherenkov telescopes array. The dark halo in the GC is usually modeled by the NFW profile \cite{Navarro:1996gj}:
\begin{equation}
\rho(r)\equiv\frac{N}{r(r-r_s)^2}\;,
\label{NFW}
\end{equation}
where $N$ is the overall normalization and $r_s$ the scale radius. This profile is in good agreement with non-baryonic cold DM simulations of the GC.
In this case and accounting for just annihilating DM, the astrophysical factor is:
$\langle J^{\text{NFW}}_{(2)} \rangle\simeq 280 \cdot 10^{23}\; \text{GeV}^2 \text{cm}^{-5}$, value that we will use as standard reference.

\section{Single-channel fits}
\label{fit}

As commented before, the particle model part of the differential gamma-ray flux expected from the GC is simulated by means of Monte Carlo
event generators, such as PYTHIA  \cite{pythia}. However, the fact that simulations have to be performed
for fixed DM mass implies that we cannot obtain explicit $M$ dependence for the photon spectra. In order to overcome
this limitation, the simulated spectra of each annihilation channel has been reproduced with the analytic fitting functions
provided in Ref. \cite{Ce10} in terms of the WIMP mass, by means of mass dependent parameters. The combination of such simulated spectra with
a power law background (Eq. (\ref{gen})) is finally fitted. We assume a typical experimental resolution of $15\%$ ($\Delta E/E\simeq0.15$)
and a perfect detector efficiency.


For quarks (except the top) electron and $\tau$ leptons, the most general formula needed to reproduce the behaviour of the differential number of photons in an energy range may be written as:

\begin{eqnarray}
\frac{dN}{dE}&=&\Big[\, a_{1}\text{exp}\left(-b_{1} \left(\frac{E}{M}\right)^{n_1}-b_2  \left(\frac{E}{M}\right)^{n_2} -\frac{c_{1}}{ \left(\frac{E}{M}\right)^{d_1}}+\frac{c_2}{ \left(\frac{E}{M}\right)^{d_2}}\right)  \nonumber \\
& + & q\, \left(\frac{E}{M}\right)^{1.5}\,\text{ln}\left[p\left(1- \left(\frac{E}{M}\right)\right)\right]\frac{ \left(\frac{E}{M}\right)^2-2 \left(\frac{E}{M}\right)+2}{ \left(\frac{E}{M}\right)}\Big]E^{-1.5}M^{0.5}\;.
\label{qs}
\end{eqnarray}

\begin{figure}[!ht]
    \centering
\resizebox{11cm}{8cm}
{\includegraphics{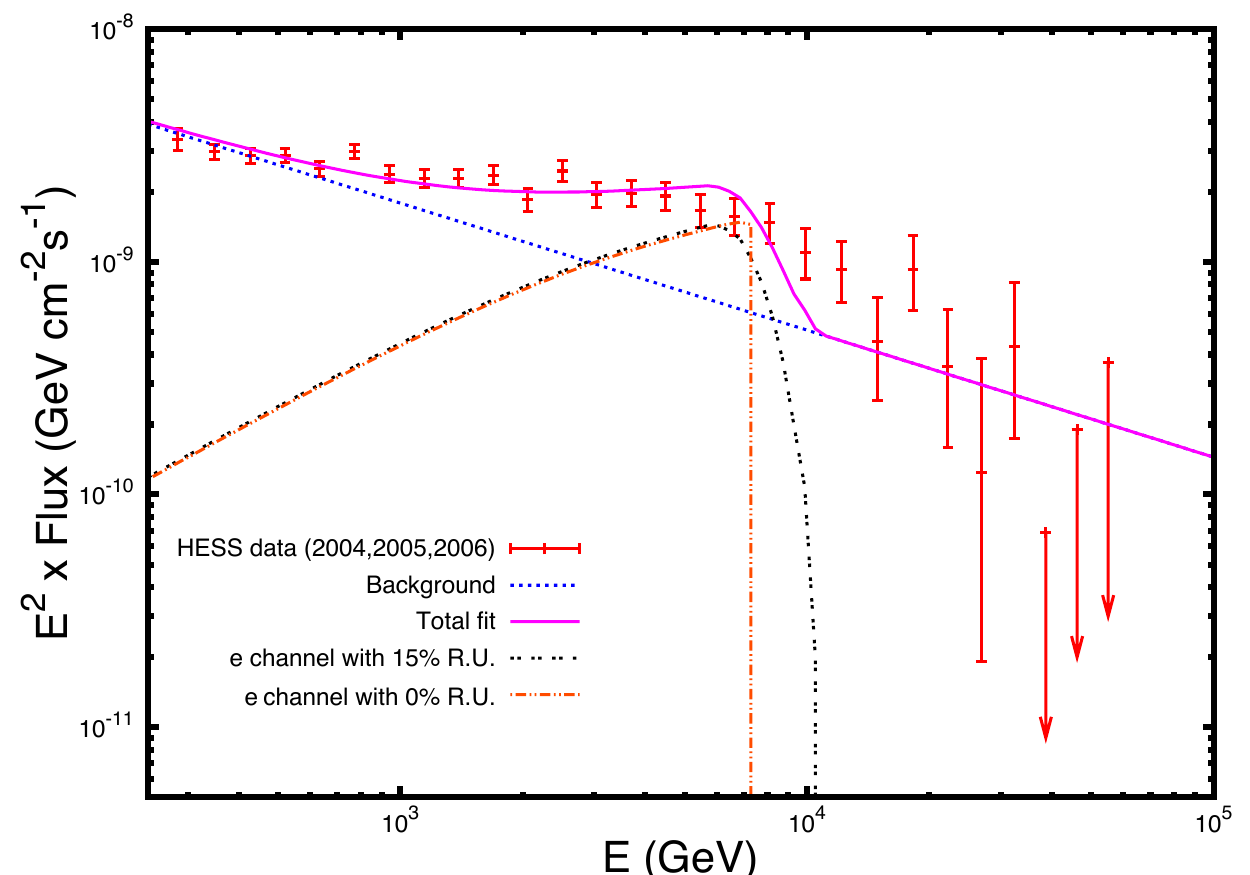}}
   \qquad
  \begin{tabular}[b]{ccccc}
  \hline
    \hline
    &Channel &     $e^+e^-$  &&      \\ \hline
    \hline
    &$M$  &  $7.51 \pm 0.11$ \\
    &$A$  & $8.12 \pm 0.73$ \\
    &$B$ & $2.78\pm0.79$  \\
    &$\Gamma$ & $2.55\pm 0.06$&\\
    &$\chi^2/\,$dof & $2.09$&\\
    &$\Delta\chi^2$ & $32.6$&\\
    &$b$ & $111 \pm  20$&\\
    \hline
    \hline
    &\\
    &\\
    &\\
    \hline
\hline
$e^-e^+$    & M       &  A         & $B$ &$\Gamma$ \\
\hline
\hline
M              &  1       &             &                  &                     \\
A               & -0.385    &  1        &                  &                     \\
$B$ &-0.044 & 0.512 & 1              &                     \\
$\Gamma$  & -0.070 & 0.583 & 0.991        & 1                     \\
\hline
\hline
&\\
&\\
    \end{tabular}
  \caption {\footnotesize{$M$(TeV), $A(10^{-7}\,\text{cm}^{-1}\text{s}^{-1/2})$,
    $B(10^{-4}\, \text{GeV}^{-1/2} \text{cm}^{-1} \text{s})$.
Best fit to the HESS J1745-290 collection of data (years 2004,2005, and 2006 \cite{HESS}) in the case that the DM
contribution came entirely from annihilation into $e^+e^-$.
The full line shows the total fitting function. The poor quality of the fit is evident ($\chi^2/dof=2.09$). The dotted line shows the simple power law background component with spectral index $\Gamma=2.55\pm0.06$, while the DM component with a $15\%$ of resolution uncertainity (R. U.) is given by the double-dotted line with a normalization parameter $A$. The dotted-dashed line shows the contribution of annihilating DM into electron-positron pairs without taking into account such a resolution.
On the top right table, we present a summary of the best fitting parameters.
$\Delta \chi^2$ correponds to the difference with respect to the best fit channel ($d\bar d$).
 On the bottom right table, we show the correlation matrix elements of this statistical analysis.
} }\label{efit}
\end{figure}

The value of the different parameters change with the SM particle annihilation channel and in some cases, 
with the range of the WIMP mass. The cases of interest are described below and the value of the parameters 
reported in Appendix A.
In the case of the electron-positron channel, the only contribution to the gamma-rays flux
comes from the bremsstrahlung of the final particles. Therefore, in the previous expression \eqref{qs}
the exponential contribution is absent, $q=\alpha_{\text{QED}}/\pi$, and $p=\left(M/m_{e^{-}}\right)^{2}$ .
This choice of the parameters corresponds to the well-known Weizs\"{a}cker-Williams approximation (Fig. \ref{efit}).

In the case of  the $\mu^+\mu^-$ channel, the exponential contribution in the expression above \eqref{qs} is also absent. A proper fitting function for such a channel can be written as:

\begin{eqnarray}
\frac{\text{d}N}{\text{d}E}\,=\, q\,\left(\frac{E}{M}\right)^{1.5}\,\text{ln}\left[p\left(1-\left(\frac{E}{M}\right)^{l}\right)\right]\frac{\left(\frac{E}{M}\right)^2-2\left(\frac{E}{M}\right)+2}{\left(\frac{E}{M}\right)}E^{-1.5}M^{0.5}\;.
\label{mu}
\end{eqnarray}

\begin{figure}[!ht]
    \centering
\resizebox{11cm}{8cm}
{\includegraphics{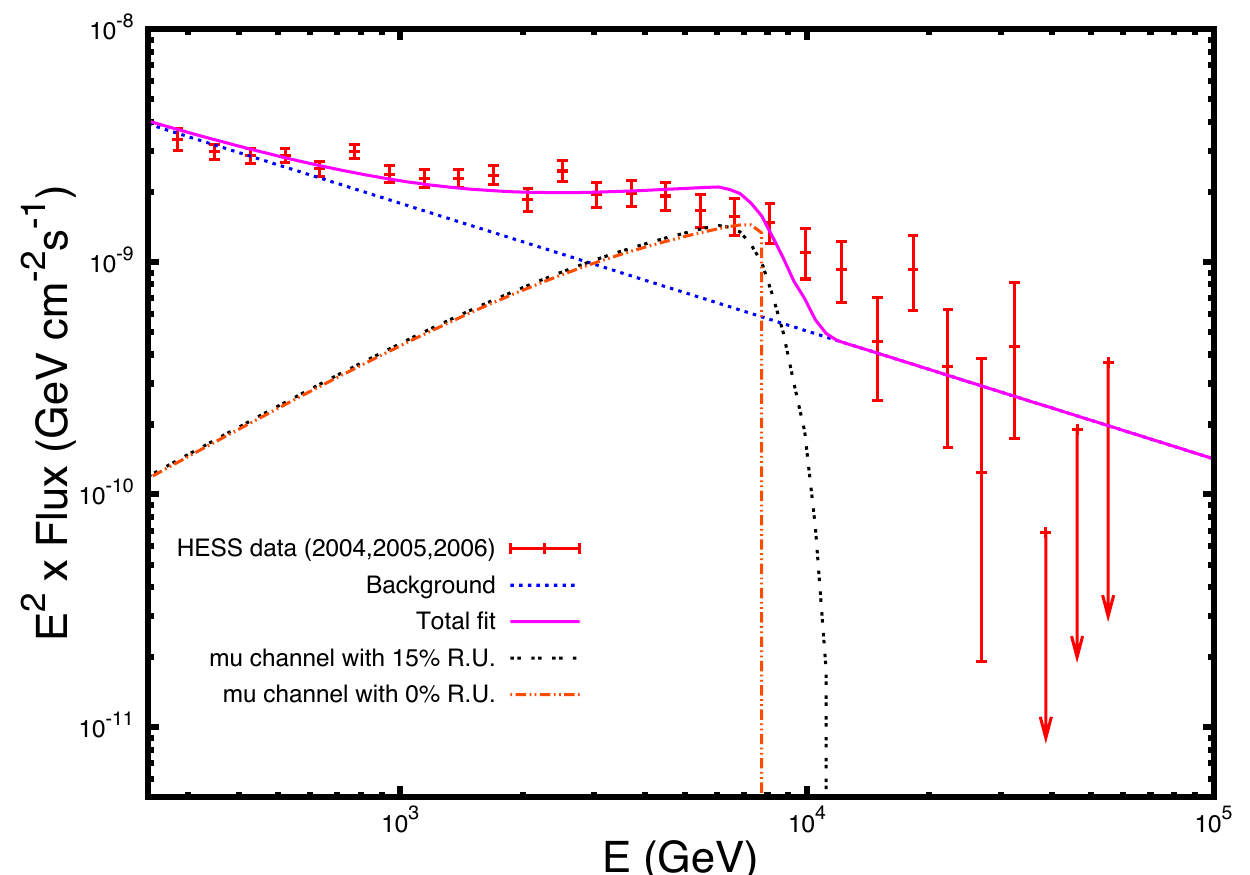}}
   \qquad
  \begin{tabular}[b]{ccccc}
  \hline
    \hline
    &Channel &     $\mu^+\mu^-$&&        \\ \hline
    \hline
    &$M$  &  $7.89 \pm 0.21$ &&\\ 
    &$A$  & $21.2 \pm 1.92$&&\\  
    &$B$ & $2.81\pm0.53$&&\\  
    &$\Gamma$ & $2.55\pm 0.06$&&\\
    &$\chi^2/\,$dof & $2.04$&&\\
    &$\Delta\chi^2$ & $31.4$&&\\
    &$b$ & $837 \pm  158$&&\\
    \hline
    \hline
    &\\
    &\\
    &\\
    \hline
\hline
$\mu^+\mu^-$    & M       &  A         & $B$ &$\Gamma$ \\
\hline
\hline
M              &  1       &             &                  &                     \\
A               & -0.431    &  1        &                  &                     \\
$B$ &-0.052 & 0.515 & 1              &                     \\
$\Gamma$  & -0.081 & 0.584 & 0.991        & 1                     \\
\hline
\hline
&\\
&\\
    \end{tabular}
\caption {\footnotesize{ The same as Fig. \ref{efit} in the case of the $\mu^+\mu^-$ annihilation channel. The fit
is only slightly better than for the $e^+e^-$ case.} }
\label{mufit}
\end{figure}

All the parameters are here mass dependent and their expression for a range of mass $10^3\,\text{GeV}\lesssim M\lesssim5\times10^4$ GeV are reported in Tab. \ref{mutabdep}. The best fit for the $\mu$ lepton is shown in Fig. \ref{mufit}.

\begin{figure}[!ht]
    \centering
\resizebox{11cm}{8cm}
{\includegraphics{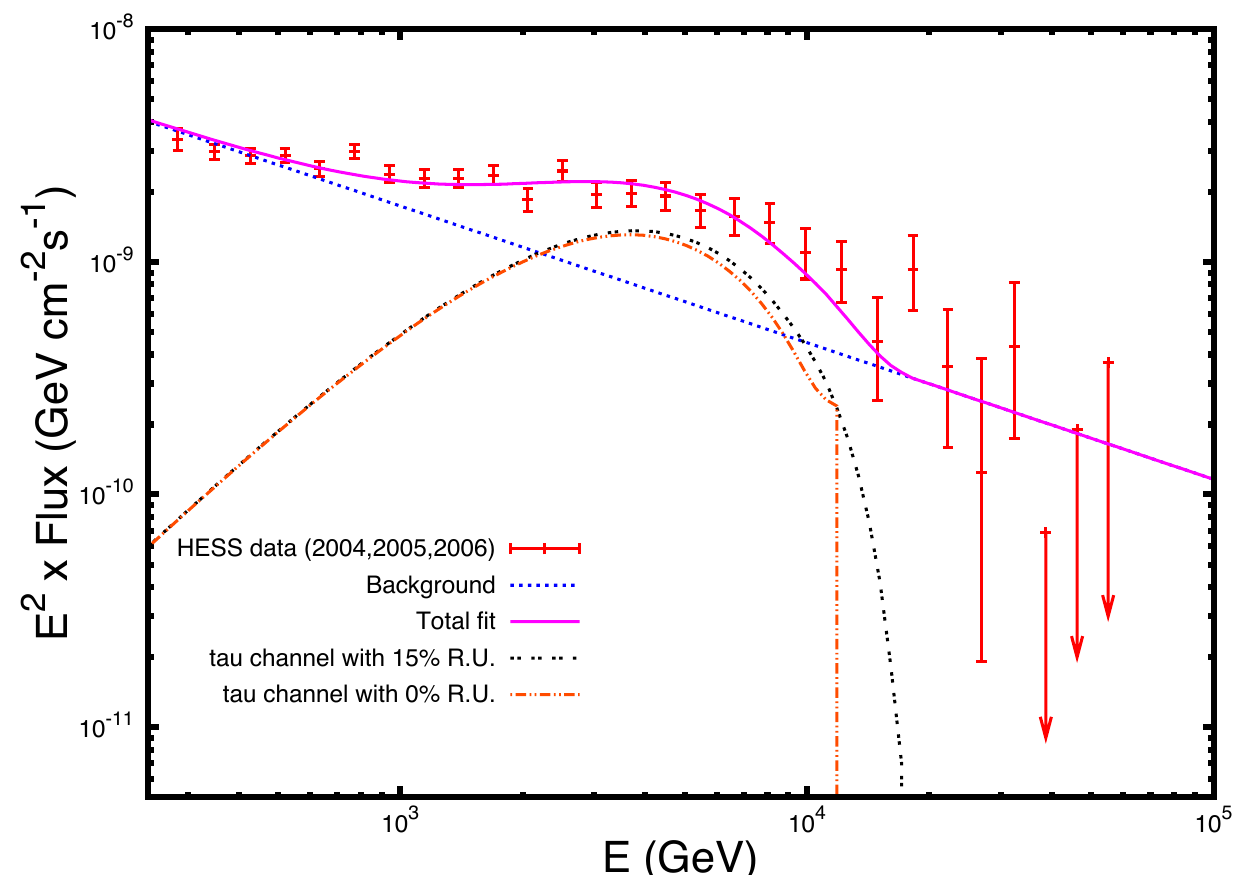}}
   \qquad
  \begin{tabular}[b]{ccccc}
  \hline
    \hline
   & Channel &     $\tau^+\tau^-$ &&        \\ \hline
    \hline
    &$M$  &  $12.4 \pm 1.3$ &&\\ 
    &$A$  & $7.78 \pm0.69$&&\\  
    &$B$ & $3.17\pm0.62$&&\\  
    &$\Gamma$ & $2.59\pm 0.06$&&\\
    &$\chi^2/\,$dof & $1.59$&&\\
    &$\Delta\chi^2$ & $20.6$&&\\
    &$b$ & $278 \pm  76$&&\\
    \hline
    \hline
    &\\
    &\\
    &\\
    \hline
\hline
   $\tau^+\tau^-$    & M       &  A         & $B$ &$\Gamma$ \\
\hline
\hline
M              &  1       &             &                  &                     \\
A               & -0.613&  1        &                  &                     \\
$B$ &0.042 & 0.487  & 1              &                     \\
$\Gamma$ & -0.004 & 0.552 & 0.993     & 1                     \\
\hline
\hline
&\\
&\\
    \end{tabular}
\caption {\footnotesize{The same as Fig. \ref{efit} in the case of the $\tau^+\tau^-$ annihilation channel.
This channel provides the best leptonic fit, although it is not enough to be
consistent with data.}}
\label{taufit}
\end{figure}

The $\tau^+\tau^-$ spectrum needs the entire Eq. (\ref{qs}) for an accurate analysis. The value of the mass independent parameter and the expression of the mass dependent ones used in this work are reported in Tab. \ref{tautab}. The best fit for the $\tau$ lepton is shown in Fig. \ref{taufit}.

\begin{figure}[!ht]
    \centering
\resizebox{11cm}{8cm}
{\includegraphics{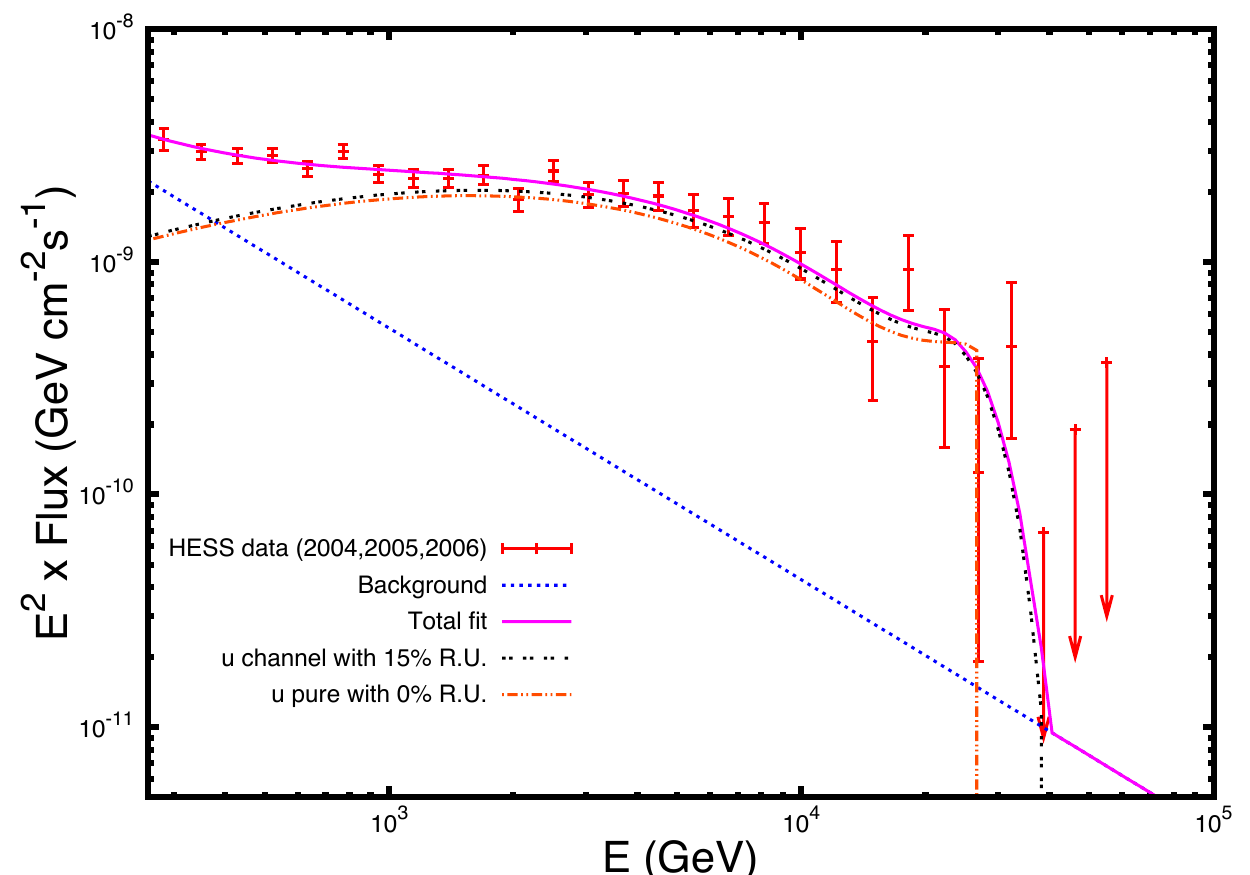}}
   \qquad
  \begin{tabular}[b]{ccccc}
  \hline
    \hline
  &  Channel &     $u\bar u$ &&       \\ \hline
    \hline
   & $M$  &  $27.9 \pm 1.8$ &&\\ 
    &$A$  & $6.51\pm0.46$&&\\  
    &$B$ & $9.52\pm9.47$&&\\  
    &$\Gamma$ & $3.08\pm 0.35$&&\\
    &$\chi^2/\,$dof & $0.78$&&\\
    &$\Delta\chi^2$ & $1.2$&&\\
    &$b$ & $987 \pm  189$&&\\
    \hline
    \hline
    &\\
    &\\
    &\\
    \hline
\hline
    $u\bar{u}$            & M       &  A         & $B$ &$\Gamma$ \\
\hline
\hline
M              &  1       &             &                  &                     \\
A               &  -0.772 &  1        &                  &                     \\
$B$ &   -0.291 & 0.768 & 1              &                     \\
$\Gamma$  & -0.315 &  0.792 & 0.999       & 1                     \\
\hline
\hline
&\\
&\\
\end{tabular}
\caption {\footnotesize{The same as Fig. \ref{efit} in the case of the $u\bar{u}$ annihilation channel.
The cut-off in the energy spectrum of this hadronic channel reproduces the cut-off in the
data, giving good consistency.} }
\label{ufit}
\end{figure}

\begin{figure}[!ht]
    \centering
\resizebox{11cm}{8cm}
{\includegraphics{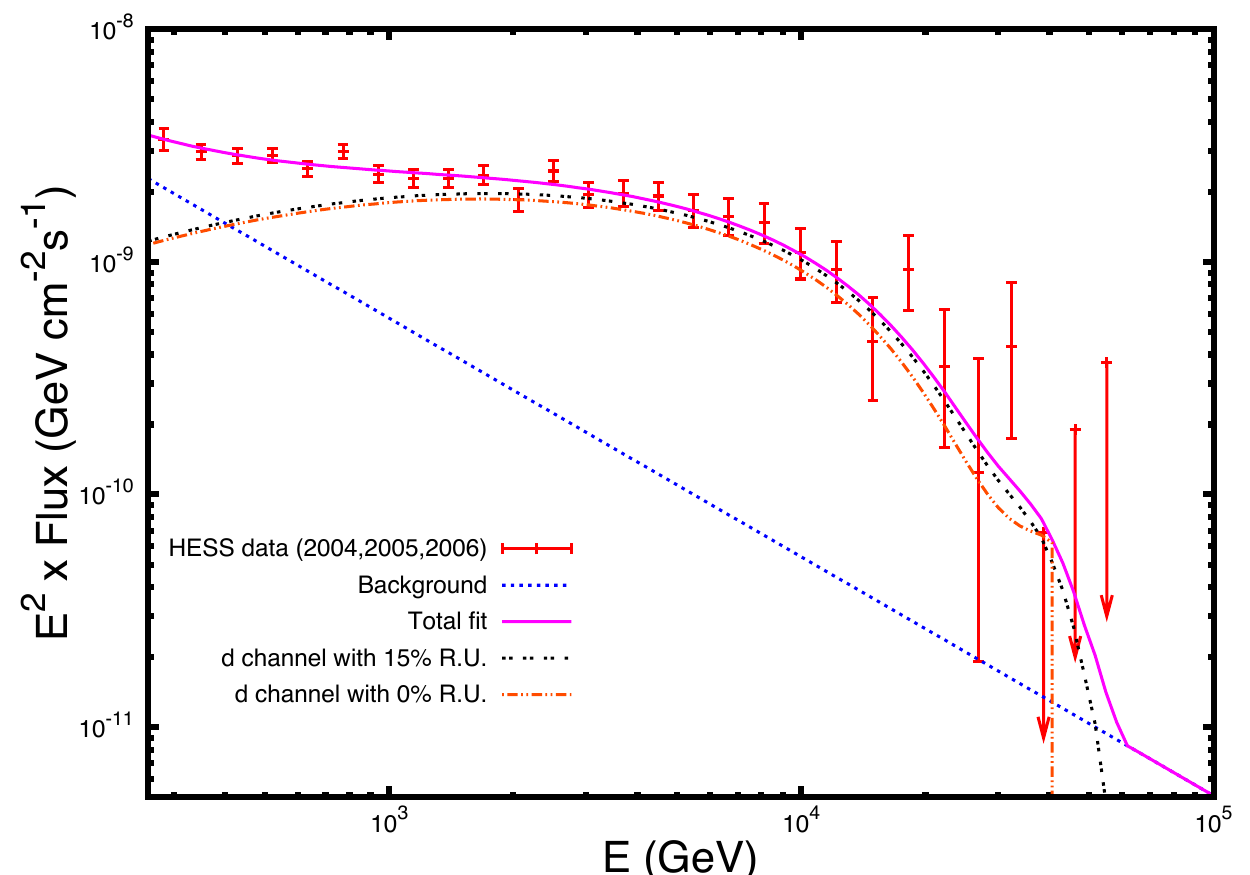}}
   \qquad
  \begin{tabular}[b]{ccccc}
  \hline
    \hline
    &Channel &     $d\bar d$   &&     \\ \hline
    \hline
    &$M$  &  $42.0 \pm 4.4$ &&\\ 
    &$A$  & $4.88\pm0.48$&&\\  
    &$B$ & $8.26\pm7.86$&&\\  
    &$\Gamma$ & $3.03\pm 0.34$&&\\
    &$\chi^2/\,$dof & $0.73$&&\\
    &$\Delta\chi^2$ & $0.0$&&\\
   & $b$ & $1257 \pm  361$&&\\
    \hline
    \hline
    &\\
    &\\
    &\\
    \hline
\hline
$d\bar d$          & M       &  A         & $B$ &$\Gamma$ \\
\hline
\hline
M              &  1       &             &                  &                     \\
A               & -0.883   &  1        &                  &                     \\
$B$ &   -0.435&  0.764 & 1              &                     \\
$\Gamma$  & -0.468 & 0.793 & 0.998        & 1                     \\
\hline
\hline
&\\
&\\
    \end{tabular}
\caption {\footnotesize{The same as Fig. \ref{efit} in the case of the $d\bar{d}$ annihilation channel.
The $\chi^2/dof$ value associated with this analysis is the lowest one obtained in this work and it has
been taken as reference to compute $\Delta\chi^2$.}}
\label{dfit}
\end{figure}

\begin{figure}[!ht]
    \centering
\resizebox{11cm}{8cm}
{\includegraphics{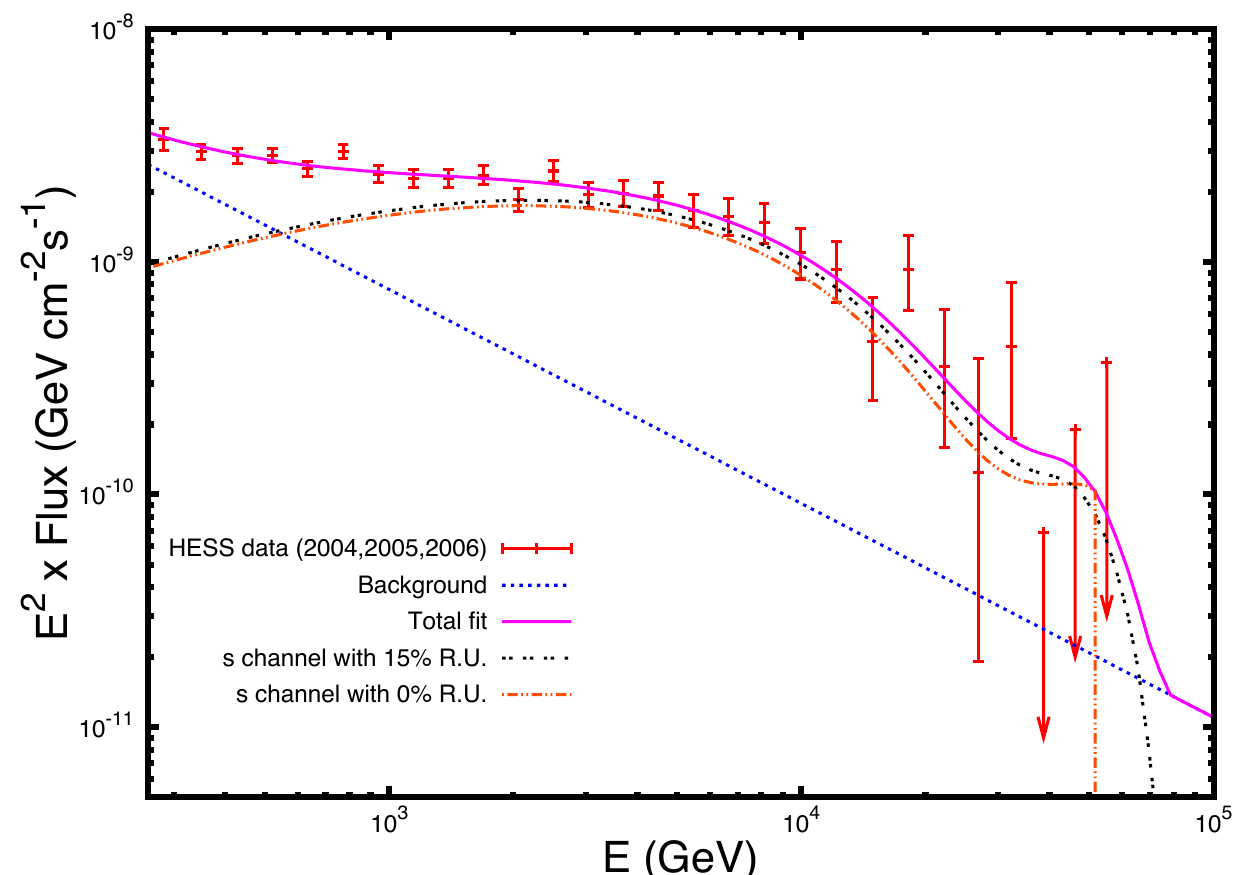}}
   \qquad
  \begin{tabular}[b]{ccccc}
  \hline
    \hline
    &Channel &     $s\bar s$ &&       \\ \hline
    \hline
    &$M$  &  $53.9 \pm 6.2$ &&\\ 
    &$A$  & $4.85\pm0.57$&&\\  
    &$B$ & $6.59\pm5.43$&&\\  
    &$\Gamma$ & $2.92\pm 0.29$&&\\
    &$\chi^2/\,$dof & $0.90$&&\\
    &$\Delta\chi^2$ & $4.1$&&\\
    &$b$ & $2045 \pm  672$&&\\
    \hline
    \hline
    &\\
    &\\
    &\\
    \hline
\hline
$s\bar s$      & M       &  A         & $B$ &$\Gamma$ \\
\hline
\hline
M              &  1       &             &                  &                     \\
A               & -0.852    &  1        &                  &                     \\
$B$ & -0.410 & 0.784 & 1              &                     \\
$\Gamma$  & -0.444 & 0.812 & 0.998         & 1                     \\
\hline
\hline
&\\
&\\
    \end{tabular}
\caption {\footnotesize{The same as Fig. \ref{efit} in the case of the $s\bar{s}$ annihilation channel.} }
\label{s fit}
\end{figure}

\begin{figure}[!ht]
    \centering
\resizebox{11cm}{8cm}
{\includegraphics{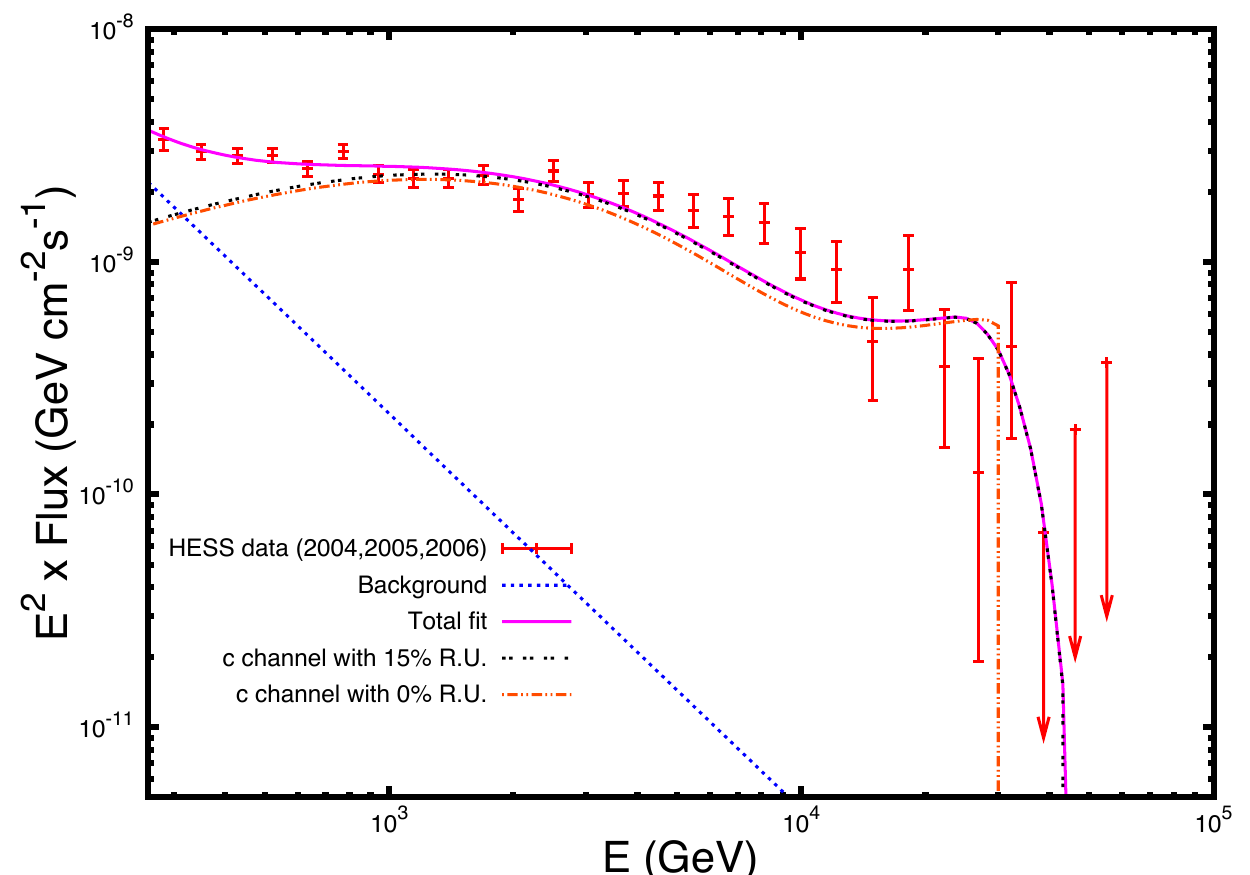}}
   \qquad
  \begin{tabular}[b]{ccccc}
  \hline
    \hline
    Channel &  &   $c\bar c$&&        \\ \hline
    \hline
    &$M$  &  $31.4 \pm 6.0$ &&\\ 
    &$A$  & $6.90\pm1.06$&&\\  
    &$B$ & $53.0\pm157$&&\\  
    &$\Gamma$ & $3.70\pm 1.07$&&\\
    &$\chi^2/\,$dof & $1.78$&&\\
    &$\Delta\chi^2$ & $25.0$&&\\
   & $b$ & $1404 \pm  689$&&\\
    \hline
    \hline
    &\\
    &\\
    &\\
    \hline
\hline
$c\bar c$           & M       &  A         & $B$ &$\Gamma$ \\
\hline
\hline
M              &  1       &             &                  &                     \\
A               &-0.970  &  1        &                  &                     \\
$B$ &  -0.738&  0.845& 1              &                     \\
$\Gamma$  & -0.754 & 0.860 & 0.999        & 1                     \\
\hline
\hline
&\\
&\\
    \end{tabular}
\caption {\footnotesize{The same as Fig. \ref{efit} but for the $c\bar{c}$ channel. This is the quark channel most disfavored by the HESS data. } }
\label{c fit}
\end{figure}

\begin{figure}[!ht]
    \centering
\resizebox{11cm}{8cm}
{\includegraphics{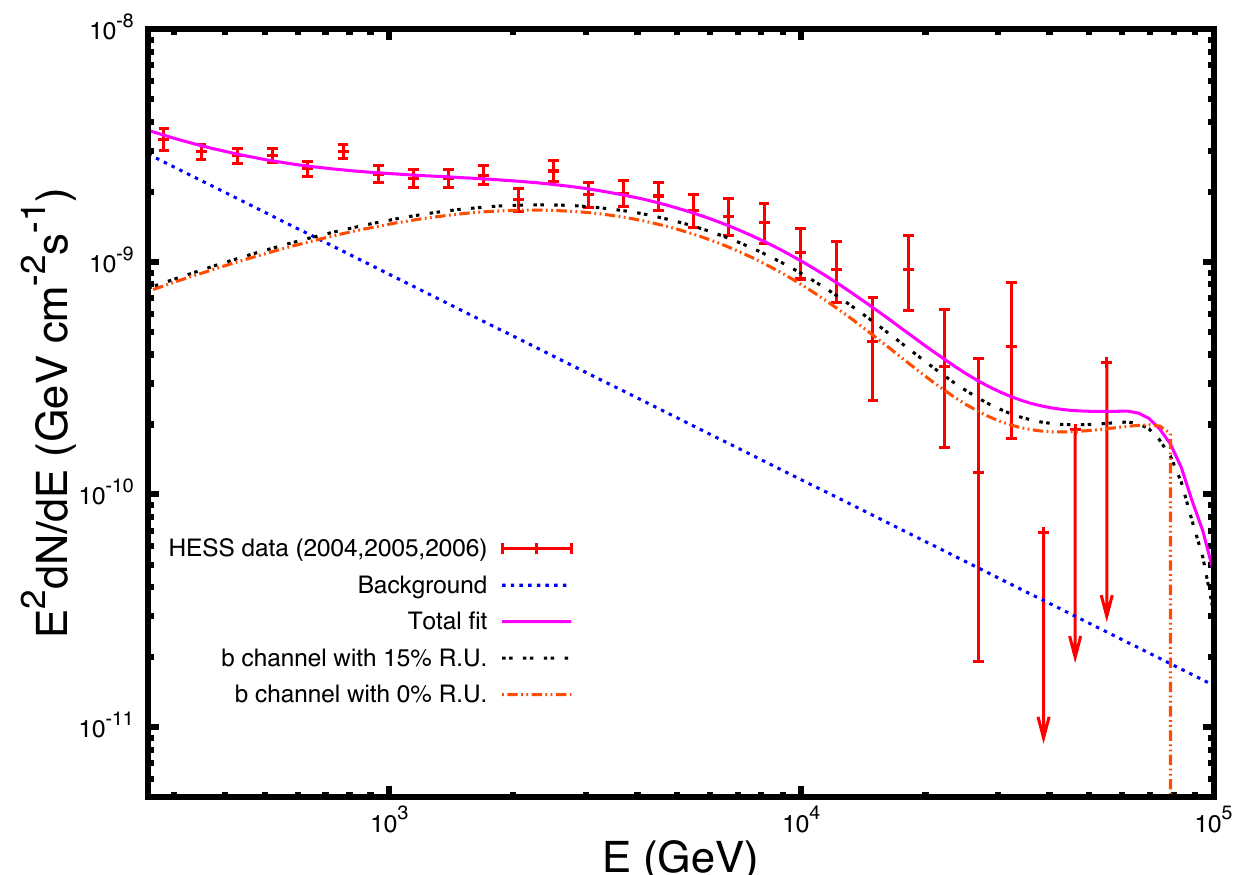}}
   \qquad
  \begin{tabular}[b]{ccccc}
  \hline
    \hline
    &Channel &     $b\bar b$&&        \\ \hline
    \hline
    &$M$  &  $82.0\pm 12.8$ &&\\ 
    &$A$  & $3.69\pm0.61$&&\\  
    &$B$ & $6.27\pm6.07$&&\\  
    &$\Gamma$ & $2.88\pm 0.35$&&\\
    &$\chi^2/\,$dof & $1.32$&&\\
    &$\Delta\chi^2$ & $14.2$&&\\
    &$b$ & $2739 \pm  1246$&&\\
    \hline
    \hline
    &\\
    &\\
    &\\
    \hline
\hline
    $b\bar b$             & M       &  A         & $B$ &$\Gamma$ \\
\hline
\hline
M              &  1       &             &                  &                     \\
A               &  -0.864  &  1        &                  &                     \\
$B$ &   -0.499 & 0.834 & 1              &                     \\
$\Gamma$  &  -0.529 & 0.857 & 0.999       & 1                     \\
\hline
\hline
&\\
&\\
    \end{tabular}
\caption {\footnotesize{The same as Fig. \ref{efit} for  the $b\bar{b}$ annihilation channel.
Together with the $c\bar{c}$ channel, is the only hadronic channel disfavored
by the HESS data.)} }

\label{b fit}
\end{figure}

The value of the mass independent parameter and the expression of the mass dependent ones in (\ref{qs}) for the $u\bar u$ channel are reported in Tab. \ref{utab}. The analytic fitting function is given for mass values up $8000$ GeV, because of limitations in the Monte Carlo event generator PYTHIA software. An extrapolation up to larger values of the mass has been performed. In this case the best fit is shown in Fig. \ref{ufit}. The parameters for the $d\bar d$ channel are given in Tab. \ref{dtab}. This channel provides the best fit (See Fig. \ref{dfit}) of all those considered in the paper and
is used as a reference for comparison with other channels. The parameters for the $s\bar s$, $c\bar c$ and $b\bar b$ are reported in Tabs. \ref{stab}, \ref{ctab} and \ref{btab}. The best fit of these hadronic channels are shown in Figs. \ref{s fit}, \ref{c fit} and \ref{b fit}.

The $t\bar t$ needs a different fitting function:

\begin{equation}
\frac{dN}{dE}=\, a_{1}\,\text{exp}\left(-b_{1}\, \left(\frac{E}{M}\right)^{n_1}-\frac{c_{1}}{\left(\frac{E}{M}\right)^{d_1}}-\frac{c_{2}}{\left(\frac{E}{M}\right)^{d_2}}\right)\left\{\frac{\text{ln}\left[p\left(1-\left(\frac{E}{M}\right)^{l}\right)\right]}{\text{ln}\,p}\right\}^{q}E^{-1.5}M^{0.5}\;.
\label{t}
\end{equation}

\begin{figure}[!ht]
    \centering
\resizebox{11cm}{8cm}
{\includegraphics{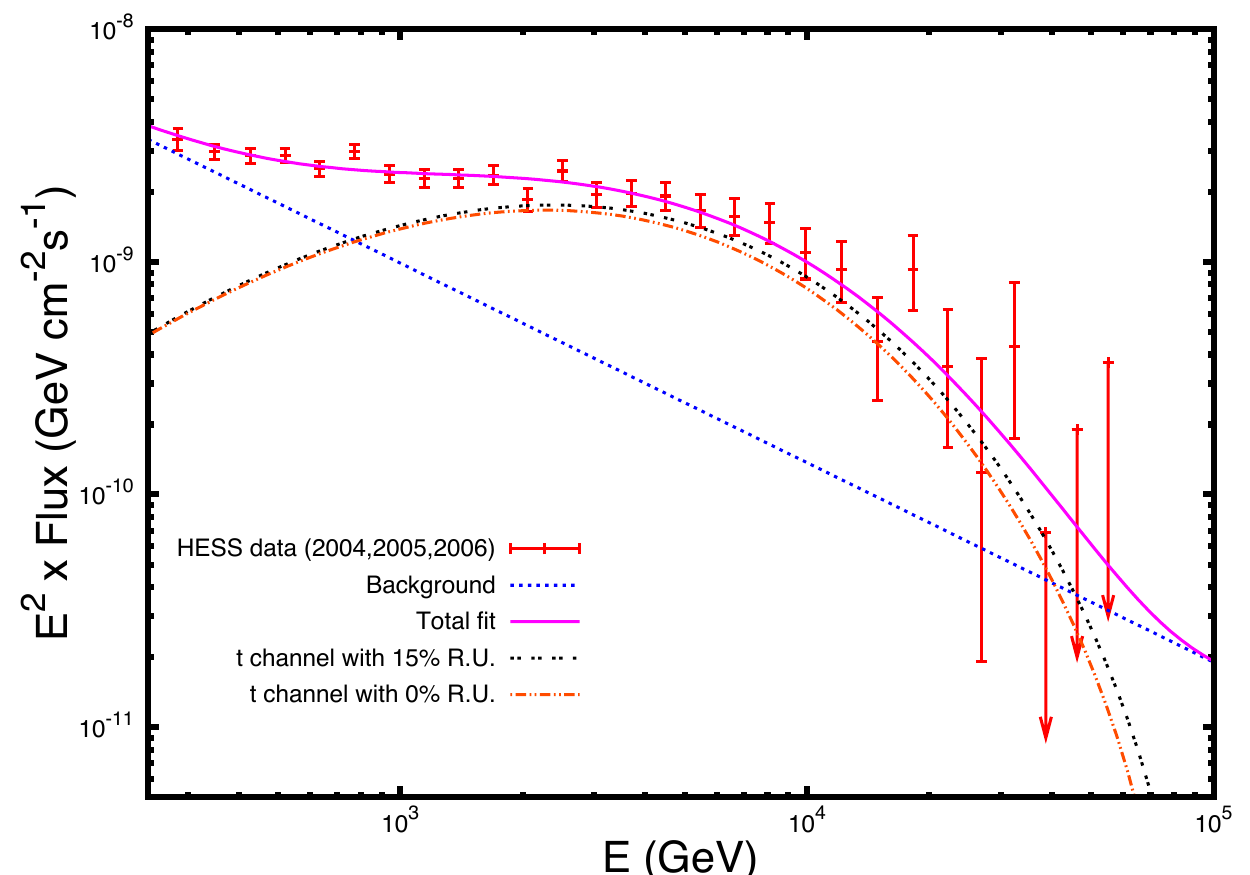}}
   \qquad
  \begin{tabular}[b]{ccccc}
  \hline
    \hline
   & Channel &     $t\bar t$ &&       \\ \hline
    \hline
  &  $M$  &  $87.7\pm 8.2$ &&\\ 
   & $A$  & $3.68\pm0.34$&&\\  
   & $B$ & $6.07\pm3.34$&&\\  
    &$\Gamma$ & $2.86\pm 0.19$&&\\
    &$\chi^2/\,$dof & $0.88$&&\\
    &$\Delta\chi^2$ & $3.6$&&\\
    &$b$ & $3116 \pm  820$&&\\
    \hline
    \hline
    &\\
    &\\
    &\\
    \hline
\hline
     $t\bar t$           & M       &  A         & $B$ &$\Gamma$ \\
\hline
\hline
M              &  1       &             &                  &                     \\
A               &  -0.748  &  1        &                  &                     \\
$B$ &-0.175 & 0.721& 1              &                     \\
$\Gamma$ &  -0.216 & 0.756 & 0.998     & 1                     \\
\hline
\hline
&\\
&\\
    \end{tabular}
\caption {\footnotesize{The same as Fig. \ref{efit} in the case of the $t\bar{t}$ annihilation channel.
The qualitative features of this channel are more similar to the electroweak annihilations (see following figures)
than to the hadronic ones (as it is shown in the previous figures). The quality of the fit is also good.} }
\label{tfit}
\end{figure}

The value of the mass independent parameter and the expression of the mass dependent ones (in the selected range of mass) are reported in Tab. \ref{ttab}.
The best fit for the $t\bar t$ channel by using Eq. (\ref{t}) is shown in Fig. \ref{tfit}.

\begin{figure}[!ht]
    \centering
\resizebox{11cm}{8cm}
{\includegraphics{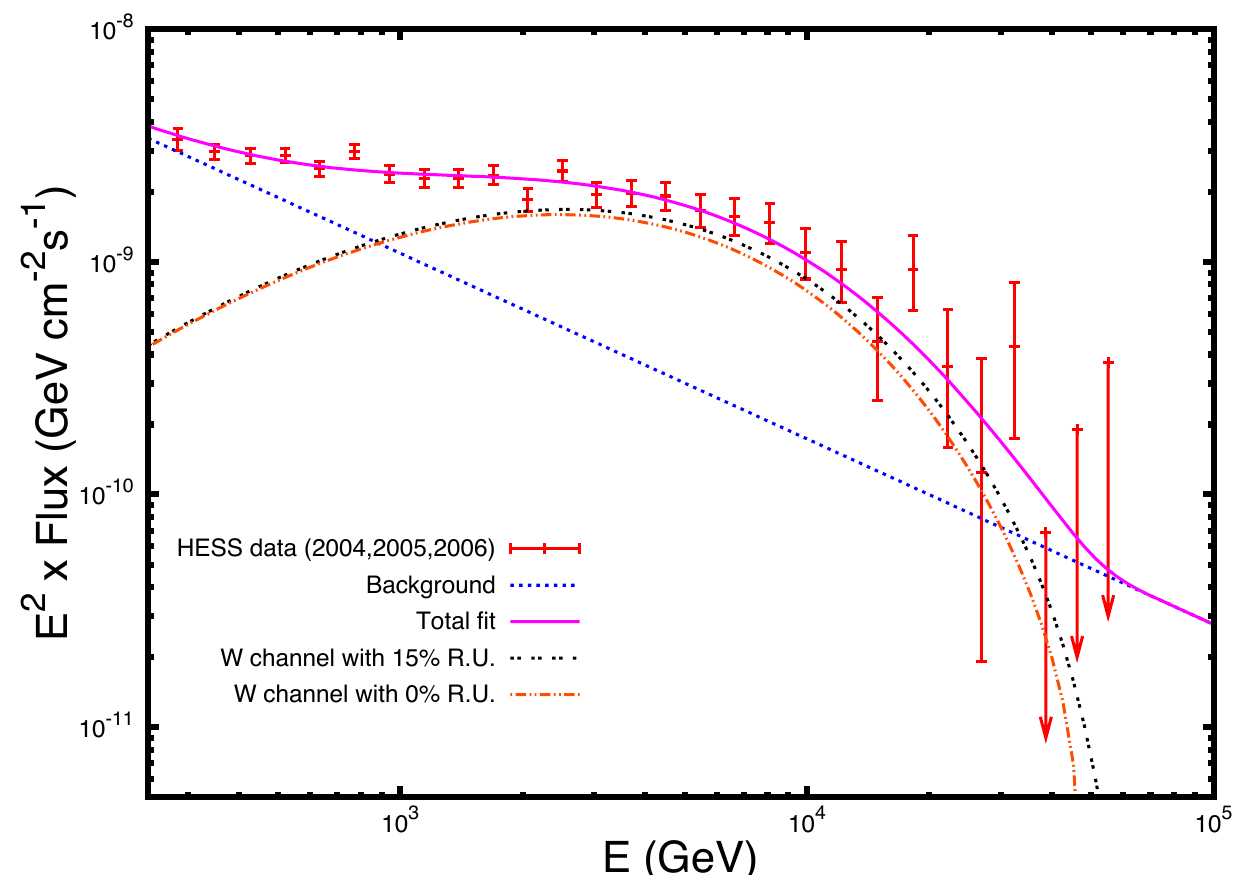}}
   \qquad
  \begin{tabular}[b]{ccccc}
  \hline
    \hline
   & Channel &     $W^+W^-$ & &       \\ \hline
    \hline
    &$M$  &  $48.8\pm 4.3$ & &\\ 
    &$A$  & $4.98\pm0.40$& &\\  
    &$B$ & $5.18\pm2.23$& &\\  
    &$\Gamma$ & $2.80\pm 0.15$& &\\
    &$\chi^2/\,$dof & $0.84$& &\\
    &$\Delta\chi^2$ & $2.6$& &\\
    &$b$ & $1767 \pm 419$& &\\
    \hline
    \hline
    &\\
    &\\
    &\\
    \hline
    \hline
     $W^+W^-$            & M       &  A         & B &$\Gamma$ \\
\hline
\hline
M              &  1       &             &                  &                     \\
A               &  -0.687&  1        &                  &                     \\
B & -0.038 & 0.681 & 1              &                     \\
$\Gamma$        & 0.008 & 0.636 & 0.997     & 1                     \\
\hline
\hline
&\\
&\\
    \end{tabular}
\caption {\footnotesize{The same as Fig. \ref{efit} for the $W^+W^-$ annihilation channel.}}
\label{Wfit}
\end{figure}
\begin{figure}[!ht]
    \centering
\resizebox{11cm}{8cm}
{\includegraphics{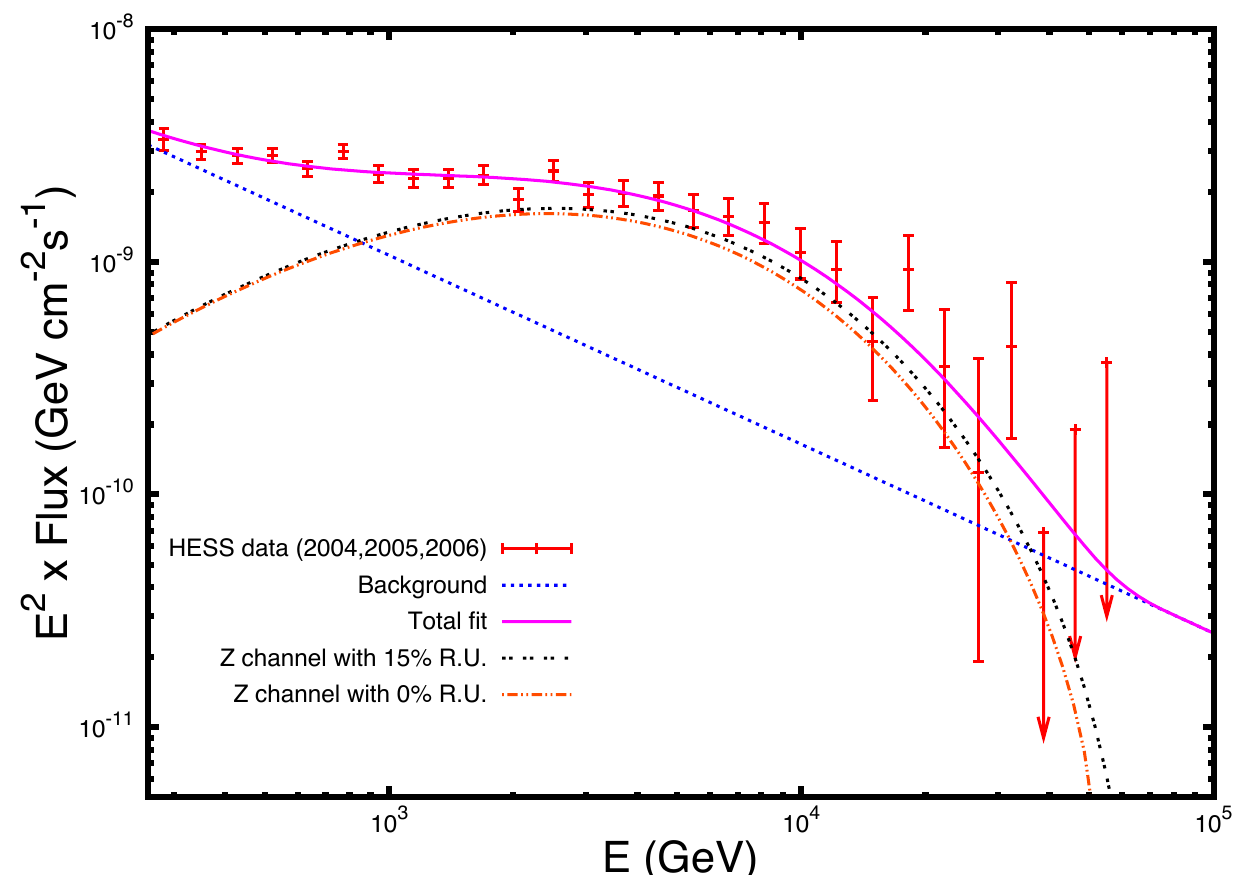}}
   \qquad
  \begin{tabular}[b]{ccccc}
  \hline
    \hline
    &Channel &     $ZZ$   & &      \\ \hline
    \hline
    &$M$  &  $54.5\pm 4.9$ & &  \\ 
    &$A$  & $4.73\pm0.40$ & & \\  
    &$B$ & $5.38\pm2.45$ & & \\  
    &$\Gamma$ & $2.81\pm 0.16$  & &\\
    &$\chi^2/\,$dof & $0.85$ & & \\
    & $\Delta\chi^2$ & $2.9$ & & \\
    & $b$ & $1988 \pm 491$ &  &\\
    \hline
    \hline
    &\\
    &\\
    &\\
\hline
\hline
     $ZZ$            & M       &  A         & B&$\Gamma$ \\
\hline
\hline
M              &  1       &             &                  &                     \\
A               &  -0.704 &  1        &                  &                     \\
B & -0.042&  0.660 & 1              &                     \\
$\Gamma$        & -0.088 &  0.703 & 0.997     & 1                     \\
\hline
\hline
&&&&\\
&&&&\\
\end{tabular}
\caption {\footnotesize{The same as Fig. \ref{efit} in the case of the $ZZ$ annihilation. The results for the $ZZ$ channel are very similar to the $W^+W^-$ one. Both electroweak channels are consistent with the data.} }
\label{Zfit}
\end{figure}

For the $W$ and $Z$ gauge bosons the parametrization is:

\begin{equation}
\frac{dN}{dE}=\, a_{1}\,\text{exp}\left(-b_{1}\, \left(\frac{E}{M}\right)^{n_1}-\frac{c_{1}}{\left(\frac{E}{M}\right)^{d_1}}\right)\left\{\frac{\text{ln}\left[p\left(j-\frac{E}{M}\right)\right]}{\text{ln}\,p}\right\}^{q}E^{-1.5}M^{0.5}\;,
\label{WZ}
\end{equation}
where the value of the parameters used in this study are reported in Tab. \ref{bos}. The best fits for the $W^+W^-$ and $ZZ$ channels are shown in Figs. \ref{Wfit} and \ref{Zfit}.


The best fit is provided by the $d\bar d$ channel with $\chi^2/dof=0.73$ for a total of 24 degrees of freedom (dof). 
In any case, other hadronic channels
such as $u\bar u$  (see Fig. \ref{ufit}) or $s\bar s$, also provide very good fits within 1$\sigma$. In the same way, softer spectra as the one
provided by $t\bar t$ (see Fig. \ref{tfit}), $W^+W^-$  (Fig. \ref{Wfit}) or $ZZ$  (Fig. \ref{Zfit}) channels are consistent with data without statistical significance difference.
On the contrary, leptonic channels (not only $e^+e^-$, or $\mu^+\mu^-$ but also  $\tau^+\tau^-$,  Figs. \ref{efit}, \ref{mufit} and \ref{taufit}), $c\bar c$ and
$b\bar b$ channels (Fig. \ref{c fit} and \ref{b fit}) are ruled out with more than 99\% confidence level
when compared to the best channel. It is interesting to note that taking into account all the channels that provide a good fit, the DM mass is constrained to
$15 \; \text{TeV} \lesssim M \lesssim 110 \; \text{TeV}$ within 2$\sigma$.  The lighter values are consistent with
hadronic annihilations ($u\bar u$) and the heavier ones with the annihilation in $t\bar t$, that is more similar to
electroweak channels.

\section{FERMI 1FGL J1745.6-2900 data}

It has been argued that the Fermi-LAT source 1FGL J1745.6-2900 and the previously considered
HESS source J1745-290 are spatially coincident \cite{Cohen}.  In \cite{ferm} data from
the first  25 months of  observations of the mentioned Fermi-LAT source have been analyzed.
It has been shown that the observed spectrum in the range 100 MeV- 300 GeV can be very well
described by a broken power law with a break energy of $E_{br}=2.0^{+0.8}_{-1.0}$ GeV, and
slopes  $\Gamma_1=2.20\pm 0.04$ and $\Gamma_2=2.68 \pm 0.05$ for lower and higher energies than
$E_{br}$ respectively. Notice that the fitted value of $\Gamma_2$ from Fermi-LAT data is in very good agreement
with the spectral index of the diffuse background obtained from HESS data in our previous analyses.
Indeed, at the  95\% confidence level, the allowed range for the spectral index
of the diffuse background is $2.6 \lesssim \Gamma \lesssim 3.7$. In this case, the lower values are consistent with all
the allowed channels, but the higher values are only accessible to the light quark channels. In any case  all  the channels that provide
a good fit to HESS data are also consistent
with the spectral index observed by Fermi-LAT.

\begin{figure}[!ht]
    \centering
\resizebox{11cm}{8cm}
{\includegraphics{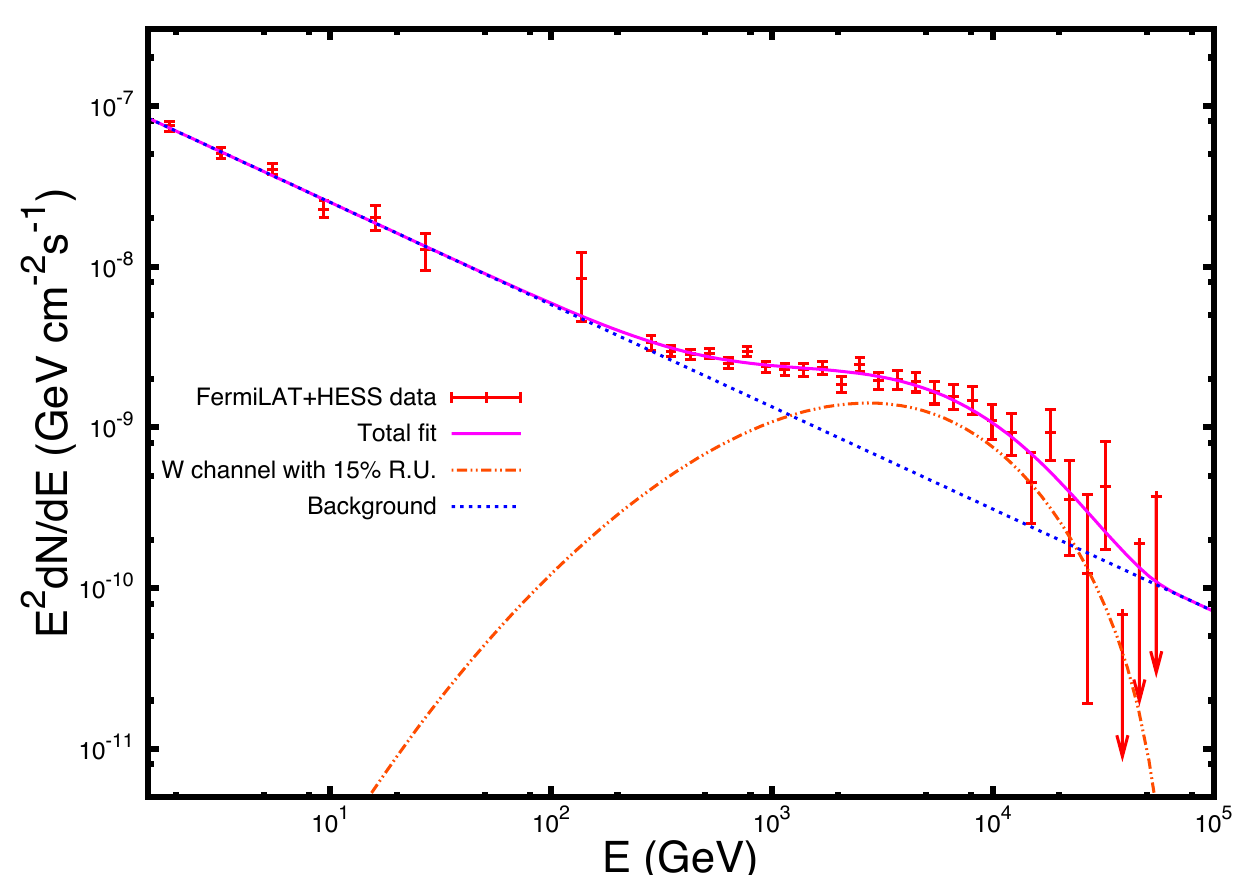}}
  \begin{tabular}[b]{ccccc}
  \hline
    \hline
   &(Fermi-LAT Data)&$W^+W^-$&\\ \hline
    \hline
    &$M$  &  $51.7\pm 5.2$ &   \\ 
    &$A$  & $4.44\pm0.34$ &  \\  
    &$B$ & $3.29\pm1.03$ &  \\  
    &$\Gamma$ & $2.63\pm 0.02$  & \\
    &$\chi^2/\,$dof & $0.75$ &  \\
    \hline
    \hline
    &\\
    &\\
    &\\
    &\\
\hline
\hline
 &$M$&$A$&$B$ &$\Gamma$ \\
 \hline
\hline
M & 1&& &   \\
A &-0.879 &1 &&\\
$B$& -0.206& 0368 & 1& \\
$\Gamma$ & -0.481& 0.718 & 0.840&1\\
\hline
\hline
&&&&\\
&&&&\\

\end{tabular}
\caption {\footnotesize{The same as in Fig. \ref{Wfit} but taking into account the Fermi-LAT data between $2 \; \text{GeV} \lesssim E \lesssim 100 \; \text{GeV}$ \cite{ferm}, that extends the dof to 27. The background parameters fitted by these data are $A=(4.44\pm0.34)\cdot10^{-7}\text{cm}^{-1}\text{s}^{-1/2}$ and
$\Gamma=2.63\pm0.02$, in agreement with the spectral index for this spectral region already found when the Fermi-LAT data are fitted with a broken power law \cite{ferm}. The quality of the fit is evident with a $\chi^2/dof=0.75$. From this interpretation of the data, Fermi-LAT instruments are sensitive just to the background component of the signal.} }
\label{WFER}
\end{figure}

In Fig. \ref{WFER}, we show the case of the $W^+W^-$ channel
to illustrate this consistency. Both the signal and background parameters are compatible with the $W^+W^-$ channel fit without the
Fermi-LAT data. With the new data, the analysis even improves to $\chi^2/dof=0.75$.
This interpretation implies that the Fermi-LAT telescope is able to detect just
the background component of the total energy spectrum of the gamma-ray emission associated with DM.

With the fit of the parameter $A$ (Eq. (\ref{A})) and by assuming a standard thermal cross section of $\langle \sigma v \rangle = 3\cdot 10^{-26}\; \text{cm}^{3} \text{s}^{-1}$, we can get an estimation of the astrophysical factor $\langle J\rangle$. We find
$10^{25}\, \text{GeV}^2\,\text{cm}^{-5}\lesssim\langle J^{\text{NFW}} \rangle\lesssim10^{26} \,\text{GeV}^2\,\text{cm}^{-5}$. It implies that the boost factors $b\equiv \langle J\rangle/\langle J^{\text{NFW}}\rangle$ spread on a range between two and three orders of magnitude.
It is interesting to note that the enhancement of the distribution of DM required to fit the data is 
compatible with the expectation of N-body simulations when the effect of the baryonic matter in the inner 
part of GC are taken into account \cite{Prada:2004pi}.

%

\section{Conclusions}
\label{con}

In this work, we have analyzed the possibility of explaining the gamma-ray data observed by HESS from the central part of our galaxy by being partially produced by DM annihilation. The complexity of the region and the ambiguous localisation of
the numerous emitting sources inside this region, seems to validate the hypothesis of a background component.
We have proved that even DM annihilations into single channels of the SM provide
good fits provided that the DM signal is complemented with a diffuse background compatible with Fermi LAT observations.
The fits returns a DM mass between $15 \; \text{TeV} \lesssim M \lesssim 110 \; \text{TeV}$ and a background spectral
index between $2.6 \lesssim \Gamma \lesssim 3.7$
within 2$\sigma$. Some channels are clearly preferred with respect to others, such as the $d\bar d$ for the quark channels with
$\chi^2/dof=0.73$, or the $W^+W^-$ and $ZZ$ channels with $\chi^2/dof=0.84$ and $\chi^2/dof=0.85$. On the contrary, the leptonic channels
are seriously disfavored. The morphology of the signal is consistent with compressed dark halos
by taking into account baryonic dissipation \cite{Blumenthal,Prada:2004pi}
with boost factors of  $b\sim 10^3$ for typical thermal cross sections.

The DM particle that may have originated these data needs to be heavier than $\sim 10$ TeV.
These large DM masses required for fitting the HESS data are not in contradiction with the unitarity
limit for thermal relic annihilation. For example, this limit is around $110$ TeV for
scalar DM particles annihilating in s-wave (for $\Omega_{\text{DM}} h^2 =0.11$) \cite{Griest:1989wd, Cembranos:2012nj}.

On the contrary, these heavy DM particles are practically unconstrained by direct detection experiments or particle colliders \cite{lab}.
An interesting example of such type of DM candidates which could have high enough mass and account for the right amount of DM in the form of a thermal relic,
is the branon \cite{branons,branonsgamma}. Branons are new degrees of freedom corresponding to brane fluctuations in brane-world models. In general, they are natural candidates for DM because they are massive fields weakly interacting with the SM particles, since their interaction is suppressed with the fourth power of the brane tension scale $f$. For masses over $100$ GeV, the main contribution to the photon spectra comes from branons annihilating into gauge bosons $ZZ$, and $W^+W^-$. In \cite{Cembranos:2012nj} it was shown that a branon DM with mass $M \simeq 50.6$ TeV,
provides an excellent fit to HESS data. The corresponding background being compatible with Fermi-LAT data. The compatibility of its thermal abundance with the WMAP constraints \cite{WMAP}, demands a
cross section of $\langle \sigma v \rangle =  (1.14\pm0.19)\cdot10^{-26}\; \text{cm}^{3} \text{s}^{-1}$, what is equivalent to a
brane tension of $f \simeq 27.5$ TeV.

The analysis of other cosmic rays \cite{cosmics} from the GC and from other astrophysical objects is fundamental to cross check the hypotheses
considered in this work. Updated analyses of this kind of signals for heavy DM combined with simple background components would be of great
interest in order to constrain the possible DM origin of the studied gamma-ray fluxes.

\vspace{0.5cm}

{\bf Acknowledgements}
We thank Alvaro de la Cruz-Dombriz for useful comments. This work has been supported by MICINN (Spain)
This work has been supported by MICINN (Spain) project numbers FIS 2008-01323, FIS2011-23000, FPA2011-27853-01
and Consolider-Ingenio MULTIDARK CSD2009-00064.


\clearpage
\section*{APPENDIX A: Fitting function parameters}
In the following tables, we show explicitly the parameters used in the fitting functions (read \cite{Ce10} for further details).

\begin{table}[h]
\begin{center}
\begin{tabular}{|c|c|}
\hline Parameter & Fitting power law(s)\\

\hline
$p$ & $4530 M^{0.653}$ \\

\hline
$q$ &  $0.00230 M^{-0.911} + 0.00291 M^{0.0348}$  \\
\hline
$l$ &  $0.626  M^{-0.0300} + 16.4  M^{-1.34}$  \\
\hline
\end{tabular}
\end{center}

\caption{Mass dependent parameters for the fitting function of the photon spectrum coming from the
$\mu^+\mu^-$ channel.}
\label{mutabdep}
\end{table}

\begin{table}[h]
\begin{center}
\begin{tabular}{|c|c|}
\hline
Parameter & Fitting power law\\
\hline
$n_{1}$ &  $-7.00 M^{-1.99}+179 M^{-0.763} + 9.09$  \\
\hline
$p$ &    $3.07 M^{1.55}$   \\
\hline
\end{tabular}\\
\begin{tabular}{|c|c|c|c|c|c|c|c|c|c|}
\hline
Channel &  $a_1$ & $b_1$ &  $b_2$ & $n_2$  & $c_1$ & $d_1$ & $c_2$& $ d_2$ & $q$ \\
\hline
$\tau$ &$14.7$ & $5.40$ & $5.31$& $1.40$ & $2.54$ &$0.295$ & $0.373$& $0.470$& $0.00260$\\
\hline
\end{tabular}\\
\end{center}
\caption{Mass dependent and independent parameters for the $\tau^+\tau^-$ channel fitting function.}
\label{tautab}
\end{table}

\begin{table}[h]
\begin{center}
\begin{tabular}{|c|c|}
\hline Parameter &  Fitting power law\\
\hline
$b_{1}$ &  $2.96 M^{0.0506}$ \\
\hline
$n_{1}$ &  $2.91 M^{-0.351} + 1.90 M^{0.0172}$\\
\hline
$n_{2}$ &  $0.0587 M^{0.146} + 0.848 M^{-0.145}$\\
\hline
$d_{1}$ &  $0.317 M^{-0.0300} + 0.403 M^{-0.351}$\\
\hline
$p$ &  $4.74 M^{0.839}$ \\
\hline
\end{tabular}\\
\begin{tabular}{|c|c|c|c|c|c|}
\hline
Channel &  $a_1$ & $b_2$ &  $c_1$ & $c_2$  & $q$ \\
\hline
$u\bar u$  & $5.58$  &5.50 &0.315 &0.0 & $9.30\cdot10^{-4}$ \\
\hline
\end{tabular}
\end{center}
\caption{Parameters for the $u\bar u$ channel fitting function.}
\label{utab}
\end{table}

\begin{table}[h]
\begin{center}
\begin{tabular}{|c|c|}
\hline
Parameter & Fitting power law\\
\hline
$b_{1}$ &  $3.39 M^{0.0485}$ \\
\hline
$n_{1}$ &  $21.8 M^{-0.993} + 2.25 M^{-0.00113}$  \\
\hline
$n_{2}$ &  $0.848 M^{-0219} + 0.161 M^{0.0573}$ \\
\hline
$c_{1}$ &  $0.722 M^{-0.270} + 0.0544 M^{0.0874}$ \\
\hline
$p$ &  $0.168 M^{1.29}$ \\
\hline
\end{tabular}\\
\begin{tabular}{|c|c|c|c|c|c|}
\hline
\hline
Channel &  $a_1$ & $b_2$ &  $d_1$ & $c_2$  & $q$ \\
\hline
$d\bar d$ & $5.20$ &$5.10$ &$0.410$&$0.0260$ &$1.40\cdot10^{-4}$ \\
\hline
\end{tabular}
\end{center}
\caption{The analogous set of parameters shown in Tab. \ref{utab} but for the $d\bar d$ channel.}
\label{dtab}
\end{table}

\begin{table}[h]
\begin{center}
\begin{tabular}{|c|c|}
\hline Parameter &  Fitting power law\\
\hline
$b_{1}$ & $4.54M^{0.0339}$ \\
\hline
$n_{2}$ & $3.68 M^{-1.01} + 0.744 M^{-0.0352}$ \\
\hline
$d_{1}$ &  $0.621 M^{-0.674} +0.414 M^{-0.0588}$  \\
\hline
$p$ &   $12.8 M^{0.732}$\\
\hline
\end{tabular}\\
\begin{tabular}{|c|c|c|c|c|c|}
\hline
\hline
Channel &  $a_1$ & $n_1$ &  $b_2$ & $c_1$  & $q$ \\
\hline
$s\bar s$ & $4.83$& $2.03$ & $6.50$ & $0.335$&$q=2.40\cdot10^{-4}$\\
\hline
\end{tabular}
\end{center}
\caption{Parameters for the fitting function of the photon spectrum coming from DM annihilation in $s\bar s$ channel.}
\label{stab}
\end{table}

\begin{table}
\begin{center}
\begin{tabular}{|c|c|}
\hline Parameter & Fitting power law\\
\hline
$b_{1}$ &  $9.90 M^{-0.130}$ \\

\hline
$n_{1}$ & $4.14 M^{-0.148}$  \\

\hline
$c_{1}$  & $0.210 M^{0.0951}$ \\
\hline
$d_{1}$ &$1.50 M^{-0.632} + 0.479 M^{-0.0942}$ \\

\hline
$p$ &  $8.11 M^{0.812}$ \\

\hline
\end{tabular}\\
\begin{tabular}{|c|c|c|c|c|c|}
\hline
\hline
Channel &  $a_1$ & $b_2$ &  $n_2$ & $c_2$  & $q$ \\
\hline
$c\bar c$ & $5.58$ & $7.90$ & $0.686$ & $0.0$&  $9.00\cdot10^{-4}$\\
\hline
\end{tabular}
\end{center}
\caption{The same as in Tab. \ref{utab} for the $c\bar c$ channel.}
\label{ctab}
\end{table}

\begin{table}[h]
\begin{center}
\begin{tabular}{|c|c|}
\hline Parameter & Fitting power law\\
\hline
$b_{1}$ & $152 M^{-0.462}$ \\

\hline
$n_{1}$ &  $18.7 M^{-0.248}$ \\

\hline
$n_{2}$     & $0.707 M^{-0.0129}$ \\
\hline
$c_{1}$ & $0.328 M^{ 0.0447}$ \\

\hline

$d_{1}$ &$0.449 M^{-0.0552}$  \\

\hline
$p$ & $11.8 M^{0.641}$ \\

\hline
\hline
\end{tabular}\\
\begin{tabular}{|c|c|c|c|c|c|}
\hline
\hline
Channel &  $a_1$ & $b_2$ &  $c_2$ & $d_2$  & $q$ \\
\hline
$b\bar b$ &$10.0$ &$11.0$ &$0.0151$ &$0.550$ &$2.60\cdot10^{-4}$\\
\hline
\end{tabular}
\end{center}
\caption{Parameters for the $b\bar b$ channel fitting function.}
\label{btab}
\end{table}

\begin{table}[h]
\begin{center}
\begin{tabular}{|c|c|}
\hline Parameter & Fitting power law(s) \\
\hline
$b_{1}$ &   $16.4 M^{-0.0400}$ \\
\hline
$n_{1}$ &  $0.559M^{-0.0379}$ \\
\hline
$c_{2}$ & $8910M^{ -3.23}$ \\
\hline
$p$ &  $5.78\cdot 10^{-5}M^{1.89}$ \\
\hline
$q$ & $0.133M^{0.488}$  \\
\hline
$l$ &  $21.9M^{-0.302}$ \\
\hline
\end{tabular}\\
\begin{tabular}{|c|c|c|c|c|}
\hline
\hline
Channel &  $a_1$ & $c_1$ & $d_1$  & $d_2$ \\
\hline
$t \bar t$    & $290$  &1.61 &0.19  & 0.845 \\
\hline
\end{tabular}
\end{center}
\caption{Mass depend and independent parameter of the fitting function given by Eq. (\ref{t}) for the $t\bar t$ channel.}
\label{ttab}
\end{table}

\begin{table}[h]
\begin{center}
\begin{tabular}{|c|c|c|c|}
\hline
\hline
Channel &  $a_1$ & $n_1$ &  $q$ \\
\hline
W  & $25.8$  & 0.510 &3.00\\
\hline
Z   & $25.8$  & 0.5 & 3.87\\
\hline
\end{tabular}\\
\begin{tabular}{|c|c|c|c|c|c|}
\hline
\hline
Channel  &$b_1$ &  $c_1$ & $d_1$  & $p$ & $j$  \\
\hline
$W^+W^-$  & $9.29M^{-0.0139}$& $0.743M^{0.0331}$ & $0.265M^{-0.0137}$ & $10^5M^{-1.13}+285M^{0.0794}$& $0.943M^{0.00852}$ \\
\hline
$ZZ$   & $9.36M^{-0.00710}$&$0.765M^{0.00980}$ & $0.272M^{-0.00990}$ & $85505M^{-0.166}+0.476M^{0.984}$& $0.884M^{0.0175}$ \\
\hline
\end{tabular}
\end{center}
\caption{Parameters for the fitting functions of the electroweak gauge bosons channels.}
\label{bos}
\end{table}


\begin{thebibliography}{99}

\bibitem{SEGUE}
MAGIC collaboration, [arXiv:astro-ph/1103.0477v1] (2011).

\bibitem{FerdSp}
A. A. Abdo et al. [arXiv:astro-ph.CO/1001.4531v1] (2010).

\bibitem{CANG}
K. Tsuchiya, R. Enomoto, L. T. Ksenofontov et al. ApJ, 606, L115 (2004).

\bibitem{VER}
K. Kosak, H. M. Badran, I. H. Bond et al., ApJ, 608, L97 (2004).

\bibitem{Aha}
F. Aharonian, A. G. Akhperjanian, K.M. Aye et al. A\&A, 425, L13 (2004b).

\bibitem{HESS}
F. Aharonian, A. G. Akhperjanian, K.M. Aye et al. A\&A, 503, 817 (2009).

\bibitem{MAG}
J. Albert, E. Aliu, H. Anderhub et al., ApJ, 638, L101 (2006).

\bibitem{Vitale}
V.~Vitale, A.~Morselli and f.~t.~F.~/L.~Collaboration, arXiv:0912.3828 [astro-ph.HE].

\bibitem{ferm}
M. Chernyakova {\em et.~al.}, ApJ {\bf 726}, 60 (2011);
T.~Linden, E.~Lovegrove and S.~Profumo,  arXiv:1203.3539 [astro-ph.HE]. 

\bibitem{Horns:2004bk}
  D.~Horns,  Phys.\ Lett.\ B {\bf 607}, 225 (2005)  [Erratum-ibid.\ B {\bf 611}, 297 (2005)].  

\bibitem{Bergstrom1}
  L.~Bergstrom, T.~Bringmann, M.~Eriksson and M.~Gustafsson,  Phys.\ Rev.\ Lett.\  {\bf 94}, 131301 (2005);
  Phys.\ Rev.\ Lett.\  {\bf 95}, 241301 (2005).

\bibitem{Profumo:2005xd}
  S.~Profumo,  Phys.\ Rev.\ D {\bf 72}, 103521 (2005). 

\bibitem{DMint}
F. Aharonian {\em et.~al.}, Phys. Rev. Lett. {\bf 97}, 221102 (2006).

\bibitem{AN}
F. Aharonian and A. Neronov, ApJ {\bf 619}, 306 (2005).

\bibitem{Cembranos:2012nj}
  J.~A.~R.~Cembranos, V.~Gammaldi and A.~L.~Maroto, Phys.\ Rev.\ D {\bf 86}, 103506 (2012). 

\bibitem{Belikov:2012ty}
  A.~V.~Belikov, G.~Zaharijas and J.~Silk, Phys.\ Rev.\ D {\bf 86}, 083516 (2012). 



\bibitem{SgrA}
R. M. Crocker, M. Fatuzzo, J. R. Jokipii et al., ApJ, 622, 892 (2005).

\bibitem{X}
Q. Wang, F. Lu and E. Gotthelf, MNRAS, 367, 937 (2006);
B. Aschenbach, N. Grosso, D. Porquet, et. al., A\&A 417, 71 (2004).

\bibitem{Atoyan}
A. Atoyan and C. D. Dermer, ApJ {\bf 617}, L123 (2004).

\bibitem{Blumenthal}
G.R. Blumenthal, S.M. Faber, R. Flores, J. R. Primack, ApJ {\bf 301}, 27 (1986);
O.~Y.~Gnedin, A.~V.~Kravtsov, A.~A.~Klypin and D.~Nagai, ApJ  {\bf 616}, 16 (2004).

\bibitem{Prada:2004pi}
F.~Prada, A.~Klypin, J.~Flix Molina, M.~Mart\'inez, E.~Simonneau, Phys.\ Rev.\ Lett.\  {\bf 93}, 241301 (2004).

\bibitem{Romano}
E.~{Romano-D{\'{\i}}az}, I.~{Shlosman}, Y.~{Hoffman}, and C.~{Heller},  ApJ {\bf 685}, L105 (2008);
ApJ {\bf 702}, 1250  (2009);
A.~V. Maccio' {\em et.~al.}, arXiv:1111.5620 [astro-ph.CO].   

\bibitem{Salucci:2011ee}
  P.~Salucci, M.~I.~Wilkinson, M.~G.~Walker, G.~F.~Gilmore, E.~K.~Grebel, A.~Koch, C.~F.~Martins and R.~F.~G.~Wyse,
  arXiv:1111.1165 [astro-ph.CO].
  
\bibitem{pythia}
T. Sjostrand, S. Mrenna and P. Skands, JHEP05 (2006) 026 (LU TP 06-13, FERMILAB-PUB-06-052-CD-T) [hep-ph/0603175].

\bibitem{Navarro:1996gj}
J.~F. Navarro, C.~S. Frenk, and S.~D. White, ApJ {\bf 490}, 493 (1997).

\bibitem{Ce10}
J.~A.~R.~Cembranos, A.~de la Cruz-Dombriz, A.~Dobado, R.~Lineros and A.~L.~Maroto,
Phys.\ Rev.\  D {\bf 83}, 083507 (2011); 
AIP Conf.\ Proc.\  {\bf 1343}, 595-597 (2011);
J.\ Phys.\ Conf.\ Ser.\  {\bf 314}, 012063 (2011);
A.~de la Cruz-Dombriz and V.~Gammaldi, arXiv:1109.5027 [hep-ph];
http://teorica.fis.ucm.es/PaginaWeb/photon\_spectra.html

\bibitem{Cohen}Cohen-Tanugi, J., Pohl, M., Tibolla, O. and Nuss, E. 2009, in Proc. 31st ICRC
(Lodz), 645 (http://icrc2009.uni.lodz.pl/proc/pdf/icrc0645.pdf)

\bibitem{Griest:1989wd}
  K.~Griest and M.~Kamionkowski, Phys.\ Rev.\ Lett.\  {\bf 64}, 615 (1990). 

\bibitem{lab}
J.~Alcaraz {\it et al.}, Phys. Rev. D {\bf 67}, 075010 (2003); 
P. Achard {\it et al.}, Phys. Lett. {\bf B597}, 145 (2004); 
J.~A.~R.~Cembranos, A.~Dobado and A.~L.~Maroto, Phys. Rev. {\bf D65} 026005 (2002); 
Phys. Rev. {\bf D70}, 096001 (2004); 
Phys.\ Rev.\ D {\bf 73}, 035008 (2006); 
Phys.\ Rev.\ D {\bf 73}, 057303 (2006); 
J.\ Phys.\ A  {\bf 40}, 6631 (2007); 
J.~A.~R.~Cembranos, J.~L.~Diaz-Cruz and L.~Prado, Phys.\ Rev.\ D {\bf 84}, 083522 (2011). 

\bibitem{branons}
 A.~Dobado and A.~L.~Maroto,  Nucl.\ Phys.\ B {\bf 592}, 203 (2001); 
J.~A.~R.~Cembranos, A.~Dobado and A.~L.~Maroto,  Phys.\ Rev.\ Lett.\  {\bf 90}, 241301 (2003); 
Phys.\ Rev.\ D {\bf 68}, 103505 (2003); 
A.~L.~Maroto, Phys.\ Rev.\ D {\bf 69}, 043509 (2004); 
Phys.\ Rev.\ D {\bf 69}, 101304 (2004); 
Int. J. Mod. Phys. {\bf D13}, 2275 (2004). 
J.~A.~R.~Cembranos  {\it et al.},  JCAP {\bf 0810}, 039 (2008). 
\bibitem{branonsgamma} J. A. R. Cembranos, A. de la Cruz-Dombriz, V. Gammaldi, A.L. Maroto,  Phys. Rev. D 85, 043505 (2012). 

\bibitem{WMAP}
E.~Komatsu {\it et al.} [WMAP Collaboration],  ApJ. Suppl. 192 18 (2011). 


\bibitem{cosmics}
S.~Rudaz and F.~W.~Stecker, Astrophys.\ J.\  {\bf 325}, 16 (1988);
J.~A.~R.~Cembranos, J.~L.~Feng, A.~Rajaraman and F.~Takayama, Phys.\ Rev.\ Lett.\  {\bf 95}, 181301 (2005); 
J.~A.~R.~Cembranos, J.~L.~Feng and L.~E.~Strigari, Phys.\ Rev.\ Lett.\  {\bf 99}, 191301 (2007); 
Phys.\ Rev.\  D {\bf 75}, 036004 (2007); 
J.~A.~R.~Cembranos and L.~E.~Strigari, Phys.\ Rev.\  D {\bf 77}, 123519 (2008); 
J.~A.~R.~Cembranos, Phys.\ Rev.\ Lett.\  {\bf 102}, 141301 (2009); 
Phys.\ Rev.\  D {\bf 73}, 064029 (2006); 
T.~Bringmann and C.~Weniger, Phys.\ Dark Univ.\  {\bf 1}, 194 (2012). 

\end{thebibliography}
\end{document}